%% file: main.tex
\documentclass[sigconf,balance=false]{acmart}
\PassOptionsToPackage{table}{xcolor}
\usepackage{xcolor}
\usepackage{colortbl}  
\usepackage{popets}
\usepackage{amsmath}
\usepackage[normalem]{ulem} %

\setcopyright{popets}
\copyrightyear{YYYY}
\acmYear{YYYY}
\acmVolume{YYYY}
\acmNumber{X}
\acmDOI{XXXXXXX.XXXXXXX}
\acmISBN{}
\acmConference{Proceedings on Privacy Enhancing Technologies}
\begin{document}
\author{Alexander Gamero-Garrido}
\affiliation{%
 \institution{University of California, Davis}
 \streetaddress{}
 \city{}
 \state{}
 \country{}
 }

\author{Kicho Yu}
\affiliation{%
 \institution{University of Southern California}
 \streetaddress{}
 \city{}
 \state{}
 \country{}
 }

\author{Sumukh Vasisht Shankar}
\affiliation{%
 \institution{Yale University}
 \streetaddress{}
 \city{}
 \state{}
 \country{}
 }

\author{Sachin Kumar Singh}
\affiliation{%
 \institution{University of Utah}
 \streetaddress{}
 \city{}
 \state{}
 \country{}
 }

\author{Sindhya Balasubramanian}
\affiliation{%
 \institution{Northeastern University}
 \streetaddress{}
 \city{}
 \state{}
 \country{}
 }

\author{Alexander Wilcox}
\affiliation{%
 \institution{Northeastern University}
 \streetaddress{}
 \city{}
 \state{}
 \country{}
 }

\author{David Choffnes}
\affiliation{%
 \institution{Northeastern University}
 \streetaddress{}
 \city{}
 \state{}
 \country{}
 }

  \title{Empirically Measuring Data Localization in the EU}

\begin{abstract}
EU data localization regulations limit data transfers to non-EU countries with the GDPR. However, BGP, DNS and other Internet protocols were not designed to enforce jurisdictional constraints, so implementing data localization is challenging. 
Despite initial research on the topic, 
little is known about if or how companies currently operate their server infrastructure to comply with the regulations.
We close this knowledge gap by empirically measuring the extent 
to which servers and routers that process EU requests are located 
outside of the EU (and a handful of ``adequate'' non-EU countries). 
The key challenge is that both browser measurements (to infer relevant endpoints) 
and data-plane measurements (to infer relevant IP addresses) are needed, 
but no large-scale public infrastructure allows both. We build a novel methodology 
that combines BrightData (browser) and RIPE Atlas (data-plane) probes,
 with joint measurements from over 1,000 networks in 19 EU countries.
We find that, on average, 2.3\% of servers %
serving users in each EU country are located in non-adequate 
destination countries (1.4\% of known trackers). 
Our findings
suggest that data localization policies are largely being followed
by content providers, though there are %
exceptions. %
\end{abstract}

  \keywords{Data Localization, GDPR, Internet Measurement}

\maketitle
\input{intro.tex}

\input{method-domains.tex}
\input{method-geolocation.tex}

\input{results-overview.tex}

\input{related.tex}

{\sloppy
\scriptsize{
\bibliographystyle{ACM-Reference-Format}
\interlinepenalty=10000
\bibliography{references}}}

\input{appendix.tex}

\end{document}

%% file: intro.tex
\section{Introduction}

Regulators around the world have taken various approaches at mitigating potential harms
resulting from unfettered collection of user data at a large scale by Internet companies, 
usually for the purpose of targeted
advertising. Data localization---broadly defined as the principle that 
requires user data to be stored in the jurisdiction where the user is physically located---is 
one such 
regulatory approach and a key component of the EU's privacy landmark regulation enacted in 2018: 
The General Data Protection Regulation (GDPR). 
A central intent behind (GDPR) data localization is to prevent transfers of user 
data to jurisdictions with weaker privacy protections, as such transfers expose 
(EU) users to unlawful commercial surveillance of user data by both commercial 
advertisers and foreign government agencies, 
as was argued in the landmark \textit{Schrems} decisions \cite{schremsi,schremsii}.

In this paper, we explore the question of whether content providers serving users in Europe 
comply with data localization regulations by locating their servers in the 
EU (or a small number of third countries that the EU regulators have approved 
data transfers to~\cite{Adequacy38:online}). In answering this question, we address a key technical 
challenge. Since data localization is based on physical boundaries, three types 
of information are simultaneously needed to audit compliance with data localization, 
and no existing platform nor methodology can procure all three: 
(\textit{i}) Which domains are most frequently visited by European users, including many additional 
domains fetched automatically 
by user-loaded websites. (\textit{ii}) The location in the network of all relevant 
domains, that is, their IP addresses. (\textit{iii}) The physical location of the servers from 
which these domains are loaded. While (\textit{iii}) is the key information needed to audit 
data localization practices, in practical terms it is difficult to obtain it directly from (\textit{i}) 
domain names; 
rather, it is easier (though still challenging ~\cite{10.1145/3402413.3402415}) to obtain the physical location of IP addresses, 
a necessary resource to connect to the Internet an actual server referenced by each domain.

Our auditing framework on data localization compliance is built with 
three components, each obtaining the necessary information described above. 
First, we identify all popular domains in each EU country using queries from BrightData~\cite{BrightData}, 
a proxy service that allows us to run a headless browser and therefore fetch complete web pages
from EU-based devices. 
Second, we launch active measurements from RIPE Atlas~\cite{RipeAtlasInfo} 
to identify the IP addresses serving each domain. Third, we use RIPE IPMap,~\cite{10.1145/3402413.3402415}) 
a database that translates IP addresses to a 
physical location (IP geolocation),
to identify IP addresses 
that are potentially located in countries 
outside the EU. We confirm a subset of these as likely 
GDPR violations using active measurements and speed-of-light constraints.
We validate our method using servers with known locations,
finding no false positives.

We find that the vast majority of servers responding to requests from EU users are 
located in the EU or adequate third countries. %
Neighbors of the EU, Russia and Turkey, are the most commonly observed destinations 
in non-adequate countries, besides the US. These servers
primarily serve requests in Finland and Romania, respectively.
We find smaller numbers of servers and known trackers in other non-adequate countries
further away from the EU, primarily in Asia.
News websites are the leading cause of these potential GDPR violations,
as they are the most common type of site
that loads trackers in non-adequate countries.
We find evidence of tracking activity taking place in a non-adequate 
country using cookies as well.
We also find significant differences across EU regions: Southern and Eastern EU
countries are more likely to be served content by known trackers located in 
non-adequate countries when compared to Western and Northern EU countries.

Combined, these findings suggest that Internet companies are likely inclined to host content 
in close proximity of users, which leads to by-default compliance with data localization 
regulations; however, there are exceptions where rigid technical constraints
may force Internet operators to serve EU users from non-compliant jurisdictions. 
In these circumstances, the imposition of physical constraints on the Internet infrastructure
is likely to be more challenging than currently recognized by both Internet operators and regulators alike,
which is especially relevant as proposed regulations on data localization emerge in more 
countries~\cite{Indiasd23:online,Dataloca93:online}. 

\section{Policy Background}
\label{sec:policy}
\if 0
 I imagine this would focus on 
 
 (1) the language of the GDPR requirement, 
 
 (2) any relevant process/case law/interpetation/legal guidelines for deeming a country to be adequate, and 
 
 (3) examples of countries that are/are not authorized for cross-border transfers, in addition to discussing the specific US 
 \fi

Data transfers to jurisdictions with weaker privacy protections
expose users in the initial jurisdiction to potential harms
such as leaks of personal information to foreign governments
or advertisers. 
Despite these potential harms, little is known about whether or how 
Internet companies that serve EU users operate their 
infrastructure to comply with this provision of the GDPR (Chapter V). 

Specifically, the GDPR in Article 45 mentions that 
``A transfer of personal data to a third country or an international organisation may take place where the [European] 
Commission has decided that the third country, a territory 
or one or more specified sectors within that third country, or the 
international organisation in question ensures an adequate level of protection.''~\cite{gdpreu}
Generally speaking, the article refers to several principles that must be followed 
in such countries receiving data transfers from EU persons, 
such as ``elements like rule of law, respect for human rights and 
fundamental freedoms, as well as whether or not data subjects' 
rights are effective and enforceable, the existence and effective 
functioning of an independent data protection authority in the non-EEA 
country and the international commitments the country or 
international organisation has entered into.''~\cite{edpb}
These adequacy decisions are published on a European Commission (EC) 
website~\cite{Adequacy38:online} and the decisions must be reviewed 
and renewed periodically by law. At the time of writing, the EC has 
determined that transfers between the EU and other European Economic Area countries 
(Norway, Liechtenstein, and Iceland) are treated as intra-EU 
transfers without needing any further safeguard. Also, according to that 
same website, the EC ``has so far recognised Andorra, Argentina, Canada 
(commercial organisations), Faroe Islands, Guernsey, Israel, Isle of Man, 
Japan, Jersey, New Zealand, Republic of Korea, Switzerland, the United 
Kingdom under the GDPR and the LED, the United States (commercial 
organisations participating in the EU-US Data Privacy Framework) 
and Uruguay as providing adequate protection.'' 

Any countries not covered by the above list are potentially inadequate, 
and any international data transfers from the EU to them would violate Article 
45 if that country is deemed inadequate by EU courts, 
\textit{e.g.}, the Court of Justice of the European Union (CJEU) or the EC. 
Even countries previously determined to be adequate may face a reversal of that decision. 
For example, in 2020, the CJEU determined in \textit{Schrems II}~\cite{schremsii}  
that ``As a consequence of such a degree of interference with the fundamental rights of persons whose
data are transferred to [the United States, where national security 
laws limit data protections], the Court declared the Privacy Shield adequacy
Decision invalid.''~\cite{edpb2}
We note that data transfers to the US were not broadly authorized when we collected our data in Aug-Sept 2022,
as the EU-US Data Privacy Framework was not adopted until December of that year ~\cite{EUUSData30:online}.
The EU-US DPF was adopted in response to the \textit{Schrems II} decision and determined adequate by the EC in July 2023.~\cite{ecreport}

Given regular instances of noncompliance resulting in EU-imposed penalties on Internet 
companies~\cite{Asitsdat63:online,30Bigges7:online}, 
it is clear that content providers are not universally
complying with these GDPR provisions. 
Besides these anecdotal instances, 
and studies with limited relevance (see
\S~\ref{sec:related}), 
there is currently no EU-wide study of whether data localization 
principles are being complied with in practice by content providers. 

Indeed, privacy advocates have primarily relied on the \textit{possibility} 
that data may be transferred to a jurisdiction with weaker privacy protections, 
especially transatlantic transfers which might result in exposure to US government surveillance. 
While these claims have successfully 
moved the needle on public policy (holding large tech companies such as Google accountable for transatlantic
data transfers~\cite{DSBAustr4:online}) through the European courts, 
they are based on analyses of the tech companies' 
own privacy policies and manual, surface-level analysis of HTML~\cite{101Compl42:online}; 
thus, these methodologies are neither aimed at nor adequate to audit compliance of 
data localization by content providers at EU scale. 

In this paper, we focus on data transfers to countries not 
determined to be adequate at the time of measurement (see definition of adequate above). 
We refer to such countries as \emph{non-adequate}.
Thus, our study identifies \textit{potential} GDPR violations, that is the existence of a tracking server in a non-adequate country.

%% file: method-domains.tex
\begin{figure}
\vspace{-4mm}
    \centering
    \includegraphics[width=.46\textwidth]{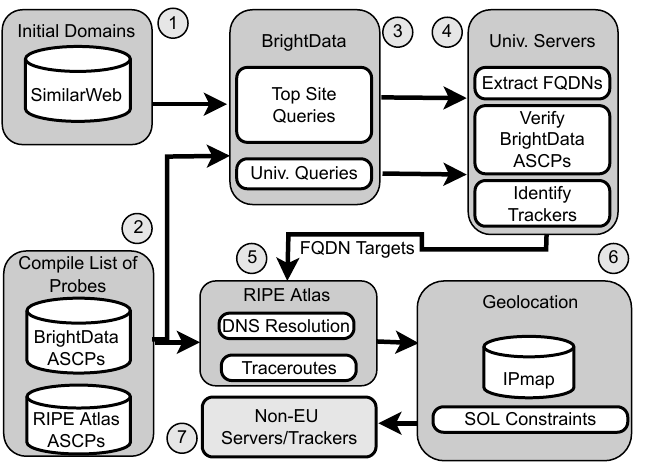}\\
    \caption{Process diagram summarizing our methodology. Steps numbered (1)\&(2) are data sources. 
Steps (3)\&(4) are browser-based measurements. Steps (5)\&(6) are data-plane measurements. 
Step (7) refers to rDNS filtering and the final output. All steps are referenced in our method sections.}
\label{fig:methodoverview}
\vspace{-3mm}
\end{figure}

\section{Network \& Country Sample}
Our method ensures that we launch both browser-based and network measurements
from the same network and in the same country. (We do not collect any personal 
information; we include an ethics section in the appendix). This constraint significantly increases the likelihood 
that both the DNS resolution, and therefore also the responding server, are 
identical in both sets of measurements. To this end, we identify overlaps in the
measurement infrastructure provided by two platforms: RIPE, the European Internet registrar
that hosts RIPE Atlas~\cite{atlas}, a large-scale Internet measurement
platform with very dense deployment in the EU; and BrightData, a large-scale proxy service~\cite{BrightData}.
To compute this overlap, we look at the networks present in each EU country that are represented
in each platform. We evaluate networks at the granularity of Autonomous Systems (AS),
the administrative domain over which a company (network operator) has control over;
ASes are the entities explicitly named in entries on the routing system, the Border Gateway Protocol, that rules
over how traffic is delivered in the global Internet. 
We thus look for AS-Country Pairs (ASCPs), or an AS in a country--a single AS can operate in 
multiple countries--where both platforms host a probe, step (2) of Fig.~\ref{fig:methodoverview}.

While RIPE regularly publishes a list of its active probes~\cite{RipeAtlasInfo},
including country and AS,
BrightData does not provide a list of active networks in each country. 
To find BrightData's AS-Country Pairs in the EU, we send repeated queries to request 
a proxy in a specific country over a period of two weeks in the last quarter of 2021. 
We find that while RIPE has presence in 2,957 ASCPs,
BrightData is present in 4,037. The intersection is 1,355 ASCPs, covering 1,318 ASes in 27 countries.

\subsection{BrightData Justification}
BrightData ensures that traffic comes from 1,000+ residential EU networks, reflecting what EU residents see on their browsers. 
Other proxy services lack this coverage. 
Further, RIPE Atlas, while widespread in the EU, lacks a browser to execute JavaScript, 
limiting its capability; data center-based approaches, meanwhile, are easily detectable and treated differently by content providers. 
Although BrightData can not load Google domains as initial sites, 
we observed 25 out of 41 Google domains (see \S~\ref{sec:googledoms}) as third parties on non-Google sites, 
capturing key interactions between browsers and Google properties.
We conduct an experiment to validate BrightData's proxy locations in \S~\ref{sec:proxyval}.

\section{Identifying Relevant Domains \& Trackers}
\label{sec:domains}
In this section, we describe our identification of relevant, popular
domains in each EU country, step (1) in Fig.~\ref{fig:methodoverview}. 
Our code and data is available on GitHub.~\cite{githubrepo}

\subsection{Initial Sample of Top Sites per EU Country}
We rely on a list of the top 50 websites in each EU country published by SimilarWeb~\cite{SimilarWeb}, 
which has also been used in previous studies~\cite{bui2022opt,zheutlin2021data}. 
SimilarWeb is the source of initial domains represented in step (1) of Fig.~\ref{fig:methodoverview}.
From this list, we exclude 19 adult sites as queries to them are not permitted by BrightData. 
 SimilarWeb has no list of top sites in 7 smaller EU countries,
so we exclude them from our sample. Due to a clerical
error in our data collection, we also exclude the Netherlands from our sample;
we describe this issue in the appendix.
We are left with 604 websites in 19 countries. 

To obtain a representative sample of popular websites in each EU country, 
we used SimilarWeb instead of Tranco,~\cite{Tranco}
as tranco primarily focuses on global rankings. 
Previous work has shown
that global website lists are skewed toward certain nations and do not reflect regional experiences.~\cite{ruth2022world}
However, to determine whether Tranco would have included 
the same regionally-popular sites, and thus would have been a reasonable alternative to SimilarWeb for our purposes, we analyzed
the percentage of top 50 websites in each EU country (based on SimilarWeb) that appeared in the Tranco list. 

\begin{center}
\begin{figure}
    \centering
    \includegraphics[width=\linewidth]{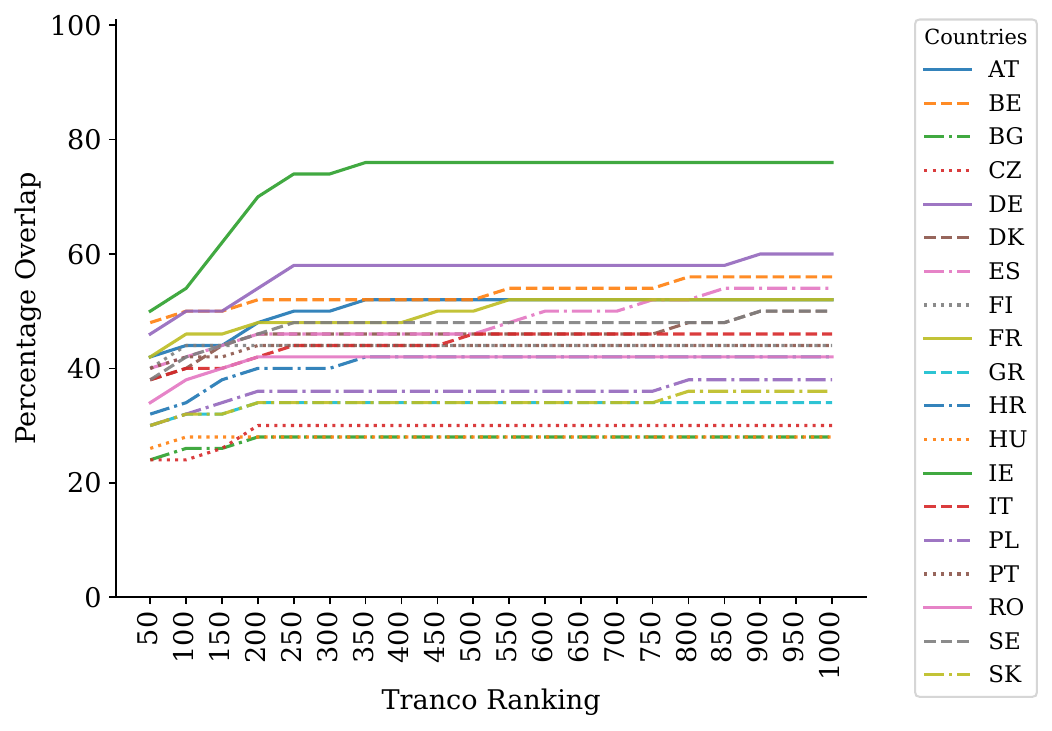}\\
    \caption{Percentage of overall agreement between Tranco and SimilarWeb lists.}
\label{fig:tranco}
\vspace{-6mm}
\end{figure}
\end{center}

Our analysis indicates that Tranco is an unsuitable replacement for SimilarWeb for country-specific website popularity. 
While Tranco's global rankings include some of the top SimilarWeb websites, the overlap is limited and inconsistent. 
Among the top 50 websites per country listed by SimilarWeb, only 24\% to 50\% appear in Tranco's top 50 (average 36\%). 
With 8 countries showing less than 35\% overlap and 14 out of 19 countries at or below 40\%. 
Even when expanding the comparison to Tranco's top 1000, overlap percentages show only modest improvement, 
ranging from 28\% to 76\% (average 45\%). With 6 countries under 40\% and 13 out of 19 at or below 50\%. 
Figure ~\ref{fig:tranco} highlights the limited overlap between SimilarWeb's country specific rankings and Tranco global rankings.

\subsection{Identification of First Parties}
\label{sec:firstparty}
To identify linked domains owned by the same entity
as the site that requests them, 
we follow a set of simple heuristics.
First, we look for AS number match derived from a sequence of
DNS resolution (in our local machine) and IP-to-AS lookup from Team Cymru~\cite{cymru}. If two domains resolve to an IP owned by 
the same AS, we infer that the domains belong to the same company and
thus the linked domain is a first party.
We similarly label domains as first parties if there is an 
organization match (using the AS from Team Cymru as an input)
in CAIDA's AS2Org database~\cite{astwoorg}.
Domains that are not first parties are then labeled as third parties.

We note that this method will produce a lower bound for 
third-party domains loaded by the initial target sites and
thus present in our data. Since our labeling of third-parties
is primarily used for finding known trackers that are missed by 
existing databases (\S~\ref{sec:trackerlabels}), 
we argue that ours is an acceptable method for labeling
domains as third parties for the scope of this paper.
We do not aim to identify every third party tracker, 
only to provide a reasonable (conservative) estimate of their prevalence,
particularly those hosted in non-adequate destinations,
in requests sent from the EU. 

\subsection{Treatment of Google Domains}
\label{sec:googledoms}
BrightData imposes restrictions on queries to Google-owned domains. 
(The reasons for these restrictions are part of a proprietary agreement
between the two companies.)
Should these queries be sent by the user, BrightData will automatically route them
through a ``superproxy,'' which is not in the ASCP that we intend to query,
deeming the results of these queries with little value for our exploration of 
data localization compliance. Thus, we are forced to exclude Google-owned properties
from our set of initial targets.

To identify Google-owned domains, we follow the first party
identification heuristics (\S~\ref{sec:firstparty}). 
Using these techniques, we identify 41 Google-owned sites
in the set of top sites from SimilarWeb. 
From these 41 sites, 28 match one of three patterns: `\url{google.TLD}',
`google.co.TLD' or `google.com.TLD', where TLD refers to any country's top-level domain, such as 
`.bg'. We also include youtube.com and news.google.com in this list, as they are well-known Google sites. 
All but 4 of these, or 24 domains, 
are present in at least one non-Google-owned site: on average, these sites
appear in more than 1,800 DNS requests from other sites--they are embedded in a vast 
number of non-Google-owned sites
in many EU countries. 
BrightData does 
allow these queries, where a Google site is loaded by a non-Google site, 
to be routed through the requested ASCP.

Of the 13 additional sites owned by Google in our set of top sites,
1 is loaded by a non-Google-owned domain. 
In sum, we are only unable to measure data localization compliance for 
16 Google-owned top sites, as the remainder are requested by non-Google sites. 
These represent less than 1\% of our target
sites from the previous subsection. 

We do acknowledge a limitation of our method (\S~\ref{sec:domexc} lists this paper's
limitations). By not directly loading the 41 Google-owned sites,
we may be missing additional trackers that target EU users. However, this limitation is mitagated
since the majority of these sites (26)
reference Google search's frontpage for a specific country, a relatively simple website that
does not typically embed a large number of non-Google sites. The limitation is further mitigated
by the fact that these frontpage Google sites are widely present in non-Google sites, so we 
are still able to infer their data localization compliance with our method (as noted
earlier, BrightData does
allow proxies to load Google sites if they are fetched by a different initial target domain).

\subsection{Final Sample of Top Sites}
To maximize response rates, we attempt to query multiple URLs for each top site.
Since a website might be responsive to only `http' or `https' requests~\cite{paracha2020deeper}, 
we attempt to query `https' first, and if we receive no response, 
we attempt `http'. Finally, we note that some sites only respond to queries
with `www.' as a prefix to the TLD+1 domain, for instance, `www.wikipedia.org' instead of
`wikipedia.org'. In sum, we attempt 4 queries for each top site, with each
subsequent query only run if the previous one failed:
\url{https://www.website.com}, \url{http://www.website.com}, \url{https://website.com}, \url{http://website.com}.

After executing our queries through BrightData
in each ASCP, we receive responses from 
534 popular sites in 20 EU countries. 

\subsection{Web Crawls Through BrightData}
\label{sec:crawling}
We ``browse'' all popular sites in each country and record their response,
step (3) of Fig.~\ref{fig:methodoverview}.
This data collection occurred in August of 2022.
Our aim is to avoid triggering anti-bot/anticrawl measures that (likely most) popular 
sites implement.
To reach this goal, we use a headless instance of Selenium with requests routed through a 
BrightData proxy: these proxies are set up on real users, and Selenium is 
a properly configured web browser (not a command-line tool such as curl).
In practical terms, we submit HTTP/HTTPS requests to each popular site in each country from
all ASCPs identified earlier. The request is sent to the target site through BrightData using 
a Python proxy handler that is initially set up for each ASCP 
with authentication information (our user ID and a plaintext passphrase),
and the proxy port.
A BrightData proxy handler follows this expression--in addition to the
previously identified fields, TCC is the two-letter country code of the requested proxy:
\begin{verbatim}
http://lum-auth-token-country-<TCC>:<passphrase>
\end{verbatim}
\begin{verbatim}
@pmgr-customer-<user\_ID>.zproxy.lum-superproxy.io:<port>
\end{verbatim}

The output of this stage is a set of DNS requests initiated by the browser, 
which executes JavaScript and other
dynamic content. These requests include the initial target site along with any 
additional domains loaded by it. These domains are the necessary information 
for our further analyses.
While the remainder of our experiments are 
based on these DNS requests, step (4) of Fig.~\ref{fig:methodoverview}, 
we also record the web contents and cookies.

\subsection{Labeling Trackers}
\label{sec:trackerlabels}
Tracking sites pose a special concern from a privacy perspective. Thus in our 
analysis we investigate compliance with data localization by all domains,
in general, and by tracking domains, in particular. 
This is depicted in step (4) of Fig.~\ref{fig:methodoverview}.
To label a domain as a tracking site, we use a three-step approach
applied to the domains found in the Selenium DNS requests (\S~\ref{sec:crawling}).
First, we intersect the domains with known trackers from the well-established list,
EasyList~\cite{EasyList}. After manually inspecting the 256 third party domains (\S~\ref{sec:firstparty}) 
we labeled as non-trackers
following this step, we found that the vast majority still appeared to be trackers. 
Thus, second, we complement EasyList with a well-known list of trackers
(with over 1k stars) on GitHub ~\cite{GitHosts}; this process yields
an additional 167 trackers. 
Third, 
for completeness, we manually inspect the remainder third party non-trackers. 
We find five additional trackers, four of which are labeled as so because of
information in their frontpage or `about us' section (\url{24media.gr}, \url{almatalent.fi}, 
\url{cdn-expressen.se}, \url{mailchimp.com}), and 
one from their WHOIS registration (labeled as `Tech Adverts', \url{amlimg.com}).
 
With our final list of tracking domains, we then manually verify 
their labeling as third-party trackers. This manual verification is necessary 
because many sites are incorrectly labeled as first parties with 
our automated approach because the initial domains 
are also hosted by a company that
itself operates a known tracker (and thus they are
delivered to users from the same network, \textit{e.g.},
the same autonomous system). For instance,
initial target site \textit{topky.sk} in Slovakia is
hosted by Google (server IP: 172.217.2.202), 
which also operates known tracker
\textit{ajax.googleapis.com}. Our automated method thus 
concludes that the latter is a first-party of the former,
when in reality they are likely not.

To verify third-party labels, we thus rely on a manual 
verification process that leverages three primary sources:
Google search of the sites involved and their ownership, 
\textit{WhoTracks.Me}~\cite{whotracksme}, a well-known database
of online trackers, and the Whois database of domain registration 
information~\cite{whois}. Given the time-intensive nature
of this verification, we restrict this analysis to the
known trackers inferred to be in non-adequate countries
by our server geolocation method (\S~\ref{sec:servergeolocation}).

%% file: method-geolocation.tex
\section{Server Geolocation}
\label{sec:servergeolocation}
This section covers our method to geolocate servers.

\subsection{Source-Based Measurements}
\label{sec:source-based}

We first obtain a preliminary assessment of where the server is located using RIPE IPmap,
step (6) of Fig.~\ref{fig:methodoverview}. %
This assessment is preliminary since even more accurate geolocation databases can err at the country level.
The passive inference %
provides us with a list of candidate server IPs that
might be located in a non-adequate country.
In this and further subsections, we aim to
identify instances of erroneous inference by IPMap;
in particular, we identify those where the server IP
is located in the EU or an adequate country but that were inferred by IPMap
as being in a non-adequate country. In other words, we look to identify false positives
in our identification of potential GDPR violations.

Our initial step to accomplish this goal
launching traceroutes towards the servers
(hostnames) previously inferred as being located in a non-adequate country,
step (5) of Fig.~\ref{fig:methodoverview}.
Note that this approach ensures that the DNS resolution is conducted 
on the same ASCP as the source of BrightData measurements.
Then, we identify candidate servers that may be located in non-adequate countries since
both the traceroute latency and IPMap support that inferred location.
This data collection took place in August of 2022.

Specifically, in this step we look for latency between the EU-based
RIPE Atlas probe and the destination server (hostname)
that is consistent with latency statistics published by Verizon~\cite{MonthlyI86:online};
this is depicted in step (6) of Fig.~\ref{fig:methodoverview}.
Since Verizon does not publish latency data between Latin America and the EU, we
rely on wondernetwork.com/pings~\cite{GlobalPi56:online} for these destinations.
In both cases, we impose a requirement that the observed latency is at least 90\%
of the average for that destination. These thresholds vary widely depending on 
the non-EU and non-adequate destination: Europe (13ms), US (65ms), EMEA region (78ms), 
Asia-Pacific region (106ms), Latin America (113-166ms depending on the country).

We launch 9,592 traceroutes towards servers in non-adequate countries (as per IPMap).\footnote{We launched 
traceroutes also towards servers in EU and adequate countries; we do not analyze these traceroutes here,
as they are unlikely to contribute information on destinations in non-adequate countries. We will
release these traceroutes with the rest of the data and code in this study.} 
In 9,012 of these cases, we analyze the traceroute latency to the last hop, 
subtracting the latency from the first hop
when possible to avoid increased latency in the last mile, \textit{e.g.}, due to WiFi.
In an additional 434 instances we use last hop latency. 
We exclude 146 traces due to either an unresponsive last hop 
(27) or latency that is higher to the first hop than the last (119).
In 8,214 traceroutes we observe latency that is below our threshold for that destination,
and we exclude these from further investigation. We are left with  1,232 traceroutes
that suggest that a server is located in a non-adequate country;
recall that the European Commission has designated a number of
non-EU member countries as being ``adequate'' for the purposes of
the GDPR's data localization requirement.~\cite{Adequacy38:online}

\subsection{Destination-Based Measurements}
\label{sec:destination-based}
To further confirm that responding servers are located in non-adequate countries, we 
collect additional evidence from RIPE Atlas probes located in those same countries.
This data collection took place in November/December of 2022.
We then use speed of light (subsequently denoted by $c$) constraints to 
discard likely erroneous geolocation inferences by IPMap.
Our goal is to remove as many false positives as possible from our experiments using empirical network data. 
However, there may still be false positives in our results; 
we describe this limitation of our work in \S~\ref{sec:falsepos}.

We launch traceroutes from RIPE Atlas probes located in the same non-adequate country
where the server was inferred to be located by IPMap, step (5) of Fig.~\ref{fig:methodoverview}. In this case, the
destination is the IP address of the server rather than a hostname,
as the DNS resolution was already done from the same network in \S~\ref{sec:source-based}.
We analyze the latency to the last hop, subtracting the latency from the first hop
as before.
We launch 598 measurements; we only measure each destination IP
once from each AS-country pair--with the AS being the 
same as that from the source-based measurements and the country 
being that inferred by IPMap for that IP--regardless of how many times the destination IP 
appears in the source-based measurements. 

We exclude 19 measurements, step (6) of Fig.~\ref{fig:methodoverview}, due to unresponsiveness of either the last hop or the 
RIPE Atlas probe, and 57 due to insufficient
granularity in the RIPE IPmap inference (or the probe's location) to compute geodesic distance. 
Of the remaining 522 measurements, in 385 cases we rely on the difference in latency between 
the last and first hops, and use the last hop latency in all others. Of these,
130 exhibit higher latency to the first hop than the last, a contradiction that we 
may be caused by additional (home) router delays due to the generation of an ICMP response,
compared to forwarding an incoming ICMP message from another device. Unlike in
\S~\ref{sec:source-based}, we keep these measurements here as by now we have
at least three pieces of evidence that the server is in a non-adequate country,
decreasing the likelihood that the server is located in the EU (recall that our goal
is to remove as many false positives as possible, as that would erroneously indicate a potential GDPR violation).
In 7 additional cases, the
latency to the first hop is not available (router did not respond to ICMP request).

We then infer whether this latency is consistent with the geodesic distance between 
the RIPE Atlas probe and the destination IP as inferred by RIPE IPmap. 
To account for the Internet's non-geodesic routing due to 
physical constraints, such as the speed of light in fiber being $2c/3$~\cite{10.1145/3402413.3402415}, 
or infrastructure delays, such as queue buildups on routers, our
upper bound for observed speed is $4c/9$~\cite{10.1145/1177080.1177090} or approx. $133 km/ms$; 
this is a more conservative 
threshold than the frequently used $2c/3$. 
If the speed inferred from the 
traceroute round-trip travel time and the geodesic distance between the endpoints 
is higher than $4c/9$, we discard the measurement,
which happens in 89 instances. We then have 433 measurements remain that target servers 
still inferred to be in non-adequate countries.

\subsection{Reverse DNS Lookups}

As a final piece of evidence in our server geolocation methods, we inspect
reverse DNS (rDNS) records of each traceroute's last hop
(reported by RIPE Atlas), step (7) of Fig.~\ref{fig:methodoverview}. Hostnames obtained from rDNS are often, 
but not always, useful in geolocating IP infrastructure~\cite{geodns}, 
which is why this is the last step in our analysis. 

Of the 433 measurements from the last subsection, 255 include
hostnames that confirm the server's country inferred in previous steps
(206 of these refer to servers in the US).
For instance, hostname \textbf{unn}-138-199-8-197.datapacket.com most likely 
refers to IP infrastructure near 
Ranong Airport (IATA code: UNN) in Thailand, which is the same country as inferred 
by IPMap for the corresponding server's IP address.
Given the diversity in operational practices to assign hostnames to IP infrastructure, 
it is not trivial to automatically
infer geographic hints to determine where the referenced infrastructure
is located; our re-implementation of recent work seemed to miss some geographic hints in hostnames~\cite{geodns}, 
which is why we manually inspect all the hostnames in this step - an 
effort that is supported by the data's manageable scale.

The rDNS records for a further 13 traceroutes suggest that the server is located in 
a different non-adequate country than that inferred by RIPE IPMap. In these cases,
we reassign the IP to the non-adequate country inferred from rDNS 
(which tends to be more accurate than latency-based inferences). 
Furthermore, the hostnames for 37 measurements suggest that the servers are located in either the EU
itself, or an adequate third-country. Nearly all of these (31) refer to AWS infrastructure 
that seems to be located in Canada but were erroneously inferred by IPMap to be in the US,
\textit{e.g.}, ec2-99-79-143-255.\textbf{ca-central-1}.compute.amazonaws.com.
We exclude these 37 IPs from further processing, as these servers are unlikely to be located in a non-adequate 
country (recall that Canada is an adequate country~\cite{Adequacy38:online}).  

Finally, 45 measurements do not return a hostname with the rDNS lookup, 
and another 83 do not seem to encode geographic locations.  We keep these servers' location 
inference unchanged from previous steps. 

\subsection{Final Sample of Non-Adequate Servers}
\label{subsec:finalsample}
We are left with 396 measurements to 247 server IP addresses where all available
evidence suggests that the server responding to EU requests is located in a non-adequate country.
These potentially non-adequate servers are present in our data in 1,233 instances,
as the same server may be contacted by multiple initial sites in each ASCP. 
Table~\ref{tab:servergeo} shows all methods we used to investigate the geolocation of servers.

\begin{table}
  \caption{Sever Geolocation. *No hostname/no geohint. **RIPE IPMap inferences.}
  \label{tab:servergeo}
  \begin{tabular}{ccccc}
    \toprule
    \textbf{Method} & \textbf{Probes} & \textbf{Unres-} & \textbf{Adequate} & \textbf{Non-}\\
     &  & \textbf{ponsive} &  & \textbf{adequate}\\
    \toprule
    Source&  9592**  &  146 &   8214 &  1232 (598 IPs)\\
    traceroutes &   &  &   & \\
    \midrule
    Destination &598 & 76 & 89 & 433\\
    traceroutes &   &  &   & \\
    \midrule
    Reverse DNS & 433 & 45/83* & 37 & 396 \\

  \bottomrule
\end{tabular}
\end{table}

%% file: results-overview.tex
\section{Analysis}
We now present our analysis of 
EU collected data. We concentrate our investigation 
along several directions. First, we investigate how frequently EU
requests are served from non-adequate destinations in each
source country.
We further analyze the prevalence of known third-party trackers
that are located in non-adequate destinations by source country.
While both of these scenarios constitute potential GDPR violations,
the latter is a stronger case since known trackers are more 
likely to be capturing personal information and 
by definition are sending such data to 
third parties.

Second, we analyze the most prevalent trackers, and the types of
websites that load them, in each source country.
Third, we investigate whether there are regional differences
across the EU in compliance with data localization.
Finally, we study the cookies loaded by sites that contact
non-adequate servers.

\subsection{Servers in Non-Adequate Countries}
\label{sec:analysis1}

We now turn to the question of prevalence of servers, generally,
and tracking servers, specifically, in non-adequate countries
by source EU country. 
Tab.~\ref{tab:countriesheat} shows a
summary of three key metrics in this regard. First, the
percentage of traceroutes sent from each EU country that reached
a server in a non-adequate country. Second, the percentage
of unique server IPs that are hosted in a non-adequate country,
and third, the same metric for unique tracking servers.

\begin{center}
\begin{figure}
    \centering
    \includegraphics[width=0.8\linewidth]{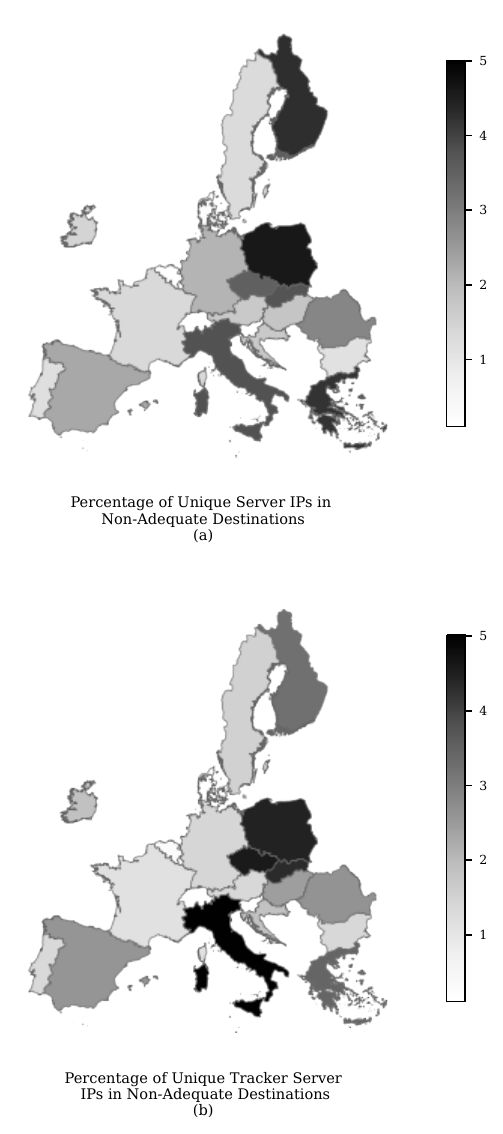}
    \caption{Percentage of unique tracker IPs in non-adequate countries.}
\label{fig:euheat}
\vspace{-6mm}
\end{figure}
\end{center}

At a high level, we find that data localization is complied with
in the vast majority of cases, as evidenced by the low
percentages in all columns of Tab.~\ref{tab:countriesheat};
we present the last two columns as a geographic heatmap in  
in Fig.~\ref{fig:euheat}.
(In \S~\ref{sec:regionalvariation}, we investigate the 
apparent geographic trends in Fig.~\ref{fig:euheat}.)
However, since
we collected data from the most popular sites in each country,
the exceptions still represent potentially very large traffic volumes
from EU countries to non-adequate countries. For instance,
4.6\% of unique IPs that serve users in Poland are located
in a non-adequate country, and the same figure is 4.2\% for Greece.
Moreover, in a majority of the 20 EU countries we studied,
more than 1\% of third-party trackers are hosted 
in non-adequate destinations. (We further investigate 
these trackers in \S~\ref{sec:trackers}.)

\begin{table}
  \caption{Percentage of traceroutes, servers and trackers reaching non-adequate destination countries, by source EU country, sorted decreasingly on row average (not shown).}
  \label{tab:countriesheat}

\begin{tabular}{cccc}
    \toprule
    \textbf{Source Country} & \textbf{Traceroutes}  & \textbf{IPs} & \textbf{Unique} \\
     & \textbf{Unique} &  & \textbf{Tracker IPs} \\
    \toprule
    RO & \cellcolor{green!100}6.6 & \cellcolor{green!42.42}2.8 & \cellcolor{green!21.21}1.4 \\
    \midrule
    FI & \cellcolor{green!65.15}4.3 & \cellcolor{green!65.15}4.3 & \cellcolor{green!16.67}1.1 \\
    \midrule
    IT & \cellcolor{green!10.61}0.7 & \cellcolor{green!57.58}3.8 & \cellcolor{green!60.61}4.0 \\
    \midrule
    GR & \cellcolor{green!21.21}1.4 & \cellcolor{green!63.64}4.2 & \cellcolor{green!43.94}2.9 \\
    \midrule
    SK & \cellcolor{green!12.12}0.8 & \cellcolor{green!56.06}3.7 & \cellcolor{green!56.06}3.7 \\
    \midrule
    PL & \cellcolor{green!12.12}0.8 & \cellcolor{green!69.70}4.6 & \cellcolor{green!36.36}2.4 \\
    \midrule
    CZ & \cellcolor{green!4.55}0.3 & \cellcolor{green!53.03}3.5 & \cellcolor{green!12.12}0.8 \\
    \midrule
    ES & \cellcolor{green!6.06}0.4 & \cellcolor{green!34.85}2.3 & \cellcolor{green!24.24}1.6 \\
    \midrule
    HR & \cellcolor{green!13.64}0.9 & \cellcolor{green!25.76}1.7 & \cellcolor{green!24.24}1.6 \\
    \midrule
    HU & \cellcolor{green!6.06}0.4 & \cellcolor{green!27.27}1.8 & \cellcolor{green!30.30}2.0 \\
    \midrule
    AT & \cellcolor{green!9.09}0.6 & \cellcolor{green!25.76}1.7 & \cellcolor{green!18.18}1.2 \\
    \midrule
    DE & \cellcolor{green!9.09}0.6 & \cellcolor{green!31.82}2.1 & \cellcolor{green!7.58}0.5 \\
    \midrule
    PT & \cellcolor{green!6.06}0.4 & \cellcolor{green!18.18}1.2 & \cellcolor{green!19.70}1.3 \\
    \midrule
    IE & \cellcolor{green!13.64}0.9 & \cellcolor{green!21.21}1.4 & \cellcolor{green!7.58}0.5 \\
    \midrule
    BG & \cellcolor{green!3.03}0.2 & \cellcolor{green!16.67}1.1 & \cellcolor{green!21.21}1.4 \\
    \midrule
    FR & \cellcolor{green!6.06}0.4 & \cellcolor{green!19.70}1.3 & \cellcolor{green!6.06}0.4 \\
    \midrule
    SE & \cellcolor{green!4.55}0.3 & \cellcolor{green!19.70}1.3 & \cellcolor{green!6.06}0.4 \\
    \midrule
    DK & \cellcolor{green!1.52}0.1 & \cellcolor{green!3.03}0.2 & \cellcolor{green!0.00}0.0 \\
    \midrule
    BE & \cellcolor{green!1.52}0.1 & \cellcolor{green!1.52}0.1 & \cellcolor{green!0.00}0.0 \\
    \bottomrule
\end{tabular}
\end{table}

\subsection{Destination Countries}
The non-adequate destination countries in our sample
span several continents. 
In Tab.~\ref{tab:traceroutesheat} we show
those most commonly occuring among these countries.
The US, Turkey and Russia account for approximately 90\% of
traceroutes crossing from the EU into a non-adequate destination.
While the US is a relatively common destination for most EU
source countries, Russia and Turkey are much more prevalent in 
two nearby source countries: Finland and Romania, respectively.

\begin{table*}
  \caption{Number of traceroutes reaching top 10 non-adequate Countries (together they account for 97.7\% of these traceroutes).}
  \label{tab:traceroutesheat}

\begin{tabular}{cccccccccccccccccccc}
    \toprule
     \textbf{Source} $\xrightarrow{}$ & \textbf{RO} & \textbf{FI} & \textbf{GR} & \textbf{HR} & \textbf{IE} & \textbf{PL} & \textbf{SK} & \textbf{IT} & \textbf{AT} & \textbf{DE} & \textbf{PT} & \textbf{FR} & \textbf{ES} & \textbf{HU} & \textbf{CZ} & \textbf{SE} & \textbf{BG} & \textbf{BE} & \textbf{DK} \\
    
    \textbf{Destination} $\downarrow$ & & & & & & & & & & & & & & & & & & \\
   
    \midrule
    \textbf{US} & \cellcolor{green!19}3 & \cellcolor{green!19}3 & \cellcolor{green!33}16 & \cellcolor{green!51}29 & \cellcolor{green!60}34 & \cellcolor{green!100}85 & \cellcolor{green!54}46 & \cellcolor{green!55}47 & \cellcolor{green!50}43 & \cellcolor{green!48}41 & \cellcolor{green!24}5 & \cellcolor{green!36}12 & \cellcolor{green!68}58 & \cellcolor{green!38}13 & \cellcolor{green!44}17 & \cellcolor{green!33}16 & \cellcolor{green!27}7 &  &  \\ 
    \midrule
    \textbf{TR} & \cellcolor{green!100}308 & \cellcolor{green!15}2 &  & \cellcolor{green!15}2 & \cellcolor{green!31}11 &  &  &  &  &  &  &  &  &  &  & \cellcolor{green!31}11 & \cellcolor{green!15}1 &  & \cellcolor{green!15}1 \\ 
    \midrule
    \textbf{RU} &  & \cellcolor{green!100}147 & \cellcolor{green!15}2 &  &  &   &  & \cellcolor{green!15}1 &  &  &  & \cellcolor{green!15}1 &  &  & \cellcolor{green!27}7 &  & \cellcolor{green!15}2 &  & \cellcolor{green!15}2 \\ 
    \midrule
    \textbf{MX} &  &  &  &  &  & \cellcolor{green!41}35 &  & \cellcolor{green!15}2 &    &  & \cellcolor{green!15}1 & \cellcolor{green!15}1 &  & \cellcolor{green!15}1 &  &  &  &  &  \\ 
    \midrule
    \textbf{IN} & \cellcolor{green!15}1 & \cellcolor{green!15}1 & \cellcolor{green!19}4 & \cellcolor{green!15}1 & \cellcolor{green!19}4 & \cellcolor{green!19}4 & \cellcolor{green!15}2 & \cellcolor{green!24}5 & \cellcolor{green!19}4 & \cellcolor{green!15}1 & \cellcolor{green!15}2 &  & \cellcolor{green!18}3 &  & \cellcolor{green!15}1 &  & \cellcolor{green!15}1 &  &  \\ 
    \midrule
    \textbf{SG} &  &  & \cellcolor{green!27}6 &  &  & \cellcolor{green!15}1 &  &  & \cellcolor{green!15}2 & \cellcolor{green!15}1 &  & \cellcolor{green!15}1 & \cellcolor{green!15}1 &  & \cellcolor{green!15}2 &  & \  &  &  \\ 
    \midrule
    \textbf{HK} & \cellcolor{green!15}1 &  & \cellcolor{green!27}7 & \cellcolor{green!15}1 & \cellcolor{green!15}1 &  &  & \cellcolor{green!15}1 &  & \cellcolor{green!24}5 & \cellcolor{green!15}1 &  & \cellcolor{green!15}1 & \cellcolor{green!15}1 & \cellcolor{green!15}1 &  &    &  &  \\ 
    \midrule
    \textbf{BR} & \cellcolor{green!19}2 &  &  &  &  &  & \cellcolor{green!15}1 & \cellcolor{green!24}5 &  &  &  & \cellcolor{green!15}1 & \cellcolor{green!15}1 &  & \cellcolor{green!19}2 &  & \cellcolor{green!19}2 & \cellcolor{green!19}2 &  \\ 
    \midrule
    \textbf{AE} &  &  &  &  &  &  &  & \cellcolor{green!18}3 & \cellcolor{green!15}1 &  &  & \cellcolor{green!15}1 & \cellcolor{green!15}1 &  & \cellcolor{green!24}4 &    & \cellcolor{green!15}1 &  &  \\ 
    \midrule
    \textbf{AU} &  &  &  &  &  & \cellcolor{green!15}1 &  & \cellcolor{green!15}1 &  & \cellcolor{green!19}2 &  &  & \cellcolor{green!15}1 &  & \cellcolor{green!15}1 &  &   &  &  \\

    \bottomrule

\end{tabular}
\end{table*}

\subsection{Trackers}
\label{sec:trackers}
We now turn our attention to known third-party trackers. As
stated earlier, these domains are of particular concern from
a privacy standpoint because of their collection of
sensitive information from users, especially in cases
where it is transferred to non-adequate countries. 
We investigate three aspects of the trackers in non-adequate countries 
observed in our data:
the source-destination pairs of countries, the most commonly occurring trackers in 
each source EU country, and the most common types of websites that host the trackers.

\subsubsection{Source-Destination Country Pairs}
\begin{figure}
    \centering
    \includegraphics[width=\linewidth]{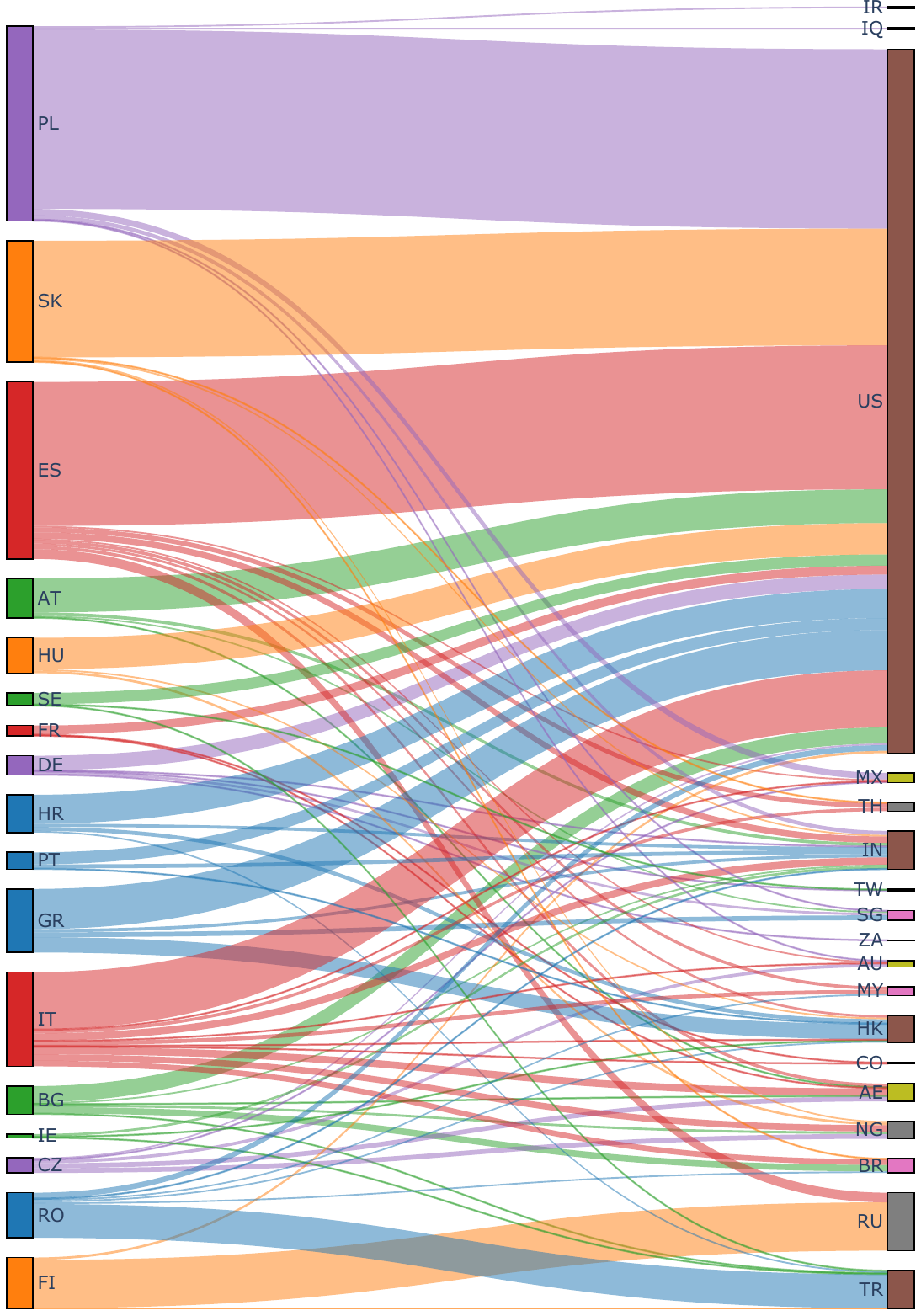}
    \caption{Flowchart showing, among 
known third-party trackers located in non-adequate countries, 
the prevalence of traceroutes connecting each source-destination pair of countries.}
\label{fig:flowsrcdst}
\end{figure}

In Fig.~\ref{fig:flowsrcdst} we show the pairs of
source EU countries and non-adequate destination country in our
data; each flow is weighted by the 
number of traceroutes seen between each pair of countries. 
As observed before for traceroutes towards all
destinations (not just trackers), the US is also the most common destination
among trackers in non-adequate countries. The US is the 
most common destination from all but four EU countries in our
sample: Ireland, Czech Republic, Romania and Finland.

Four destination countries are also notable in Fig.~\ref{fig:flowsrcdst}.
India is observed as a destination from 13 EU countries; in over
80\% of cases (among 49 traceroutes), a Google (third-party) tracker is responsible
for this. Russia, Turkey and
Hong Kong are frequent destinations from three EU countries: Finland, Romania
and Greece, respectively. Thus, the trend observed earlier for all traceroutes
from the former two EU sources holds also for known trackers. 
A majority (26/43) of the traceroutes between Romania and Turkey are
caused by a single popular website: \textit{fandom.com},
a site focused on arts and entertainment ~\cite{SimilarWeb}.
In the case of Finland and Russia, exactly half (30/60) of
the traceroutes are caused by popular website \textit{vk.com},
a social media network~\cite{SimilarWeb} that is popular among Russian speakers.
(Russian speakers are a minority group in Finland, which borders Russia).
Finally, 17 of 19 traceroutes
from Greece to Hong Kong are caused by a single third-party tracker: Facebook, 
whoose trackers are loaded by 10 popular websites. \textit{makeleio.gr}, a news and
media publisher~\cite{SimilarWeb}, is the most commonly occurring popular website, responsible for 4/17 of
these traceroutes. 

\subsection{Common Paths}

\begin{figure}
    \centering
    \includegraphics[width=\linewidth]{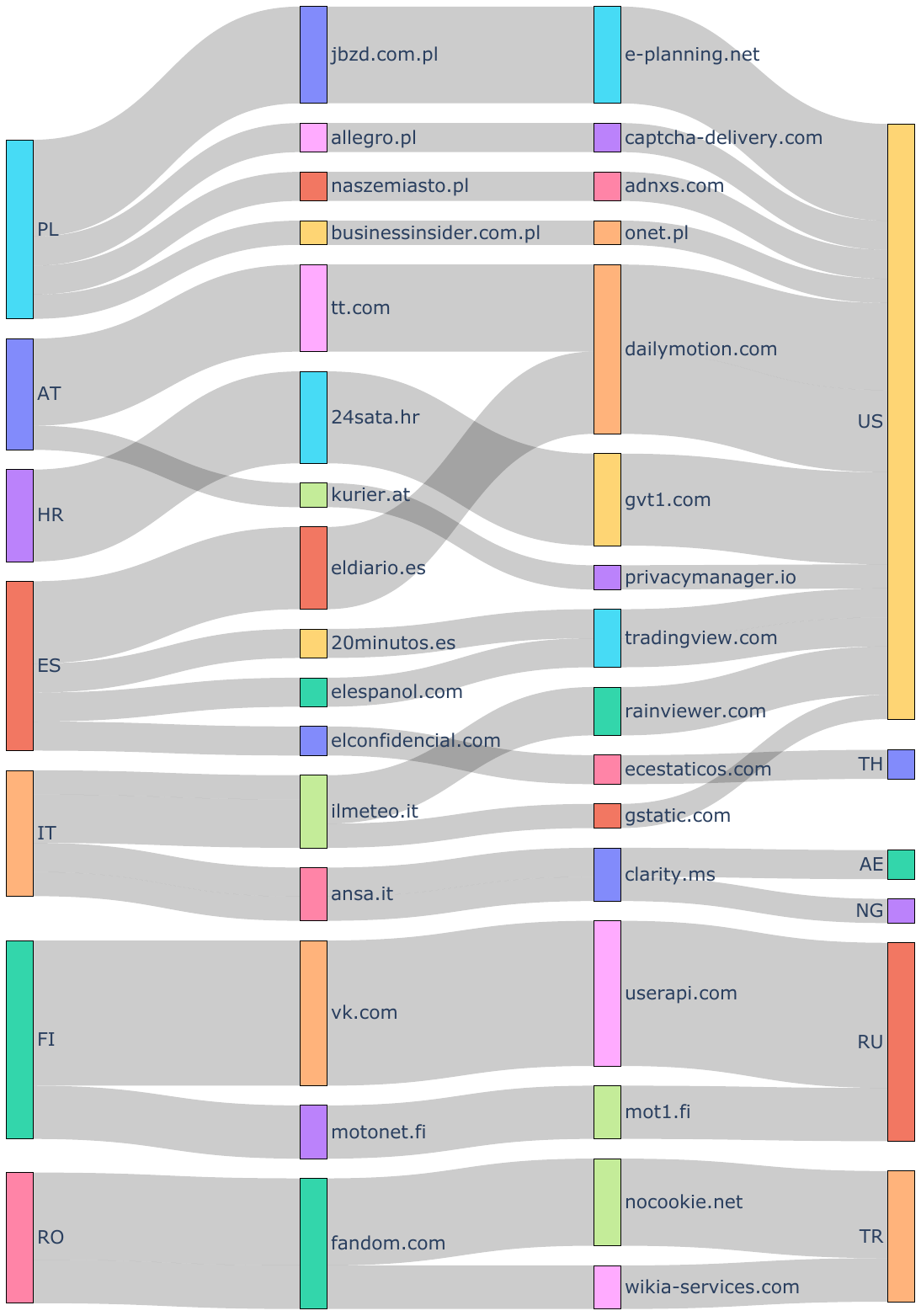}\\
    \caption{Combinations of
source country, initial website, third-party tracker, and destination country that are observed
five times or more in our data.}
\label{fig:flowcmmntrk}
\end{figure}

We now investigate common paths that take a request from an EU country
to a non-adequate destination. We define a path as a combination of 
source country, initial website, tracker, and destination country. 
Among these paths, we show those that occur at least 5 times in our traceroute
set in Fig.~\ref{fig:flowcmmntrk}. Two of these paths, which originate in 
Romania and Finland through \textit{fandom.com} and \textit{vk.com}, respectively, were mentioned earlier.
(\textit{userapi.com} might be ``affiliated'' with \textit{vk.com}, but we could not
confirm the former's ownership by the latter, thus we classify this site as a third-party.)

At a high level, we observe a variety of paths that are not particularly concentrated 
among any one website or tracker. Notably, only four of these trackers seem to be operated by major
US companies: \textit{dailymotion.com}, \textit{gvt1.com} and \textit{gstatic.com} (Google), 
and \textit{clarity.ms} (Microsoft). As before, there are a wide variety of paths that reach the US 
through combinations of popular websites and trackers.

\subsection{Initial Website Categories}
\label{sec:categories}

\begin{figure}[t]
    \centering
    \includegraphics[width=\linewidth]{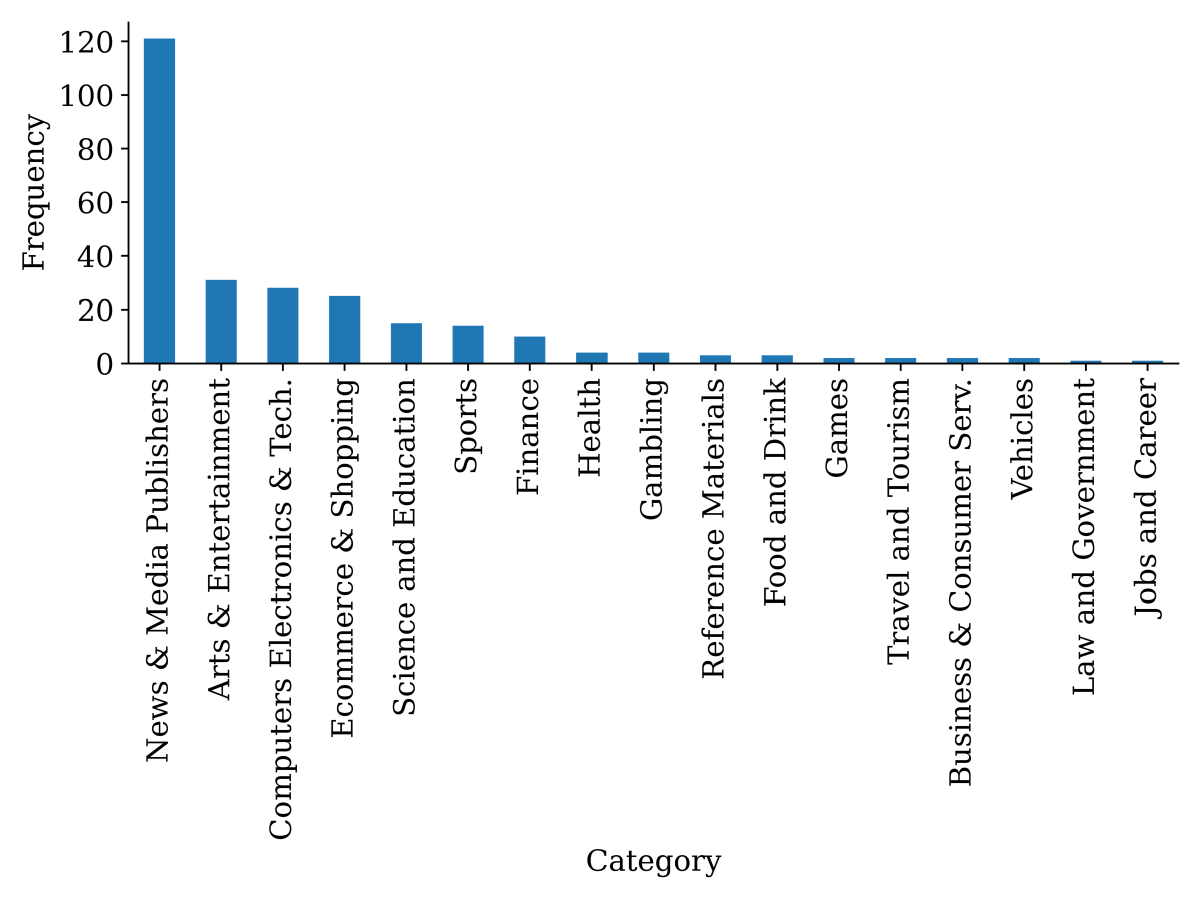}\\
    \caption{Categories for websites that load observed trackers in non-adequate destinations.}
\label{fig:categories}
\end{figure}

We now turn to the question of which types of websites
are responsible for loading these third-party trackers in
non-adequate countries. Anectodally, news sites seem prevalent among these;
for instance, in Fig.~\ref{fig:flowcmmntrk}, all the initial websites in Spain
are news websites. 
We now systematically investigate whether that is broadly the case.
In Fig.~\ref{fig:categories} we
present the \textit{SimilarWeb} categories~\cite{SimilarWeb} associated with the 239 websites
that load at least one of these trackers. A slim majority of these sites, 120, 
belong to the News \& Media Publishers category (including the four aforementioned sites in Spain). 
This category makes intuitive sense
as a frequent fetcher of trackers due to the news industry's increasing reliance
on digital advertising revenue and thus on web tracking. Nevertheless, the high
prevalence of this category in the set of sites loading trackers in non-adequate countries is still notable.
For instance, three other categories include at least 20 websites: Arts \& Entertainment, Computers Electronics and Technology,
and Ecommerce \& Shopping. However, none of them come close to the prevalence of News \& Media Publishers.

\subsection{Regional Variation}
\label{sec:regionalvariation}
In previous subsections, especially in 
Fig.~\ref{fig:euheat}, we anecdotally observed that the rates of 
servers located in non-adequate countries seemed higher in Southern and
Eastern Europe. To evaluate whether this is true, systematically,
we use the United Nations defition~\cite{uneu} for four regions of Europe: 
Northern, Southern, Eastern and Western.
We evaluate both the rate of server IPs, overall, and tracker IPs, specifically.

\begin{figure}
    \centering
    \includegraphics[width=\linewidth]{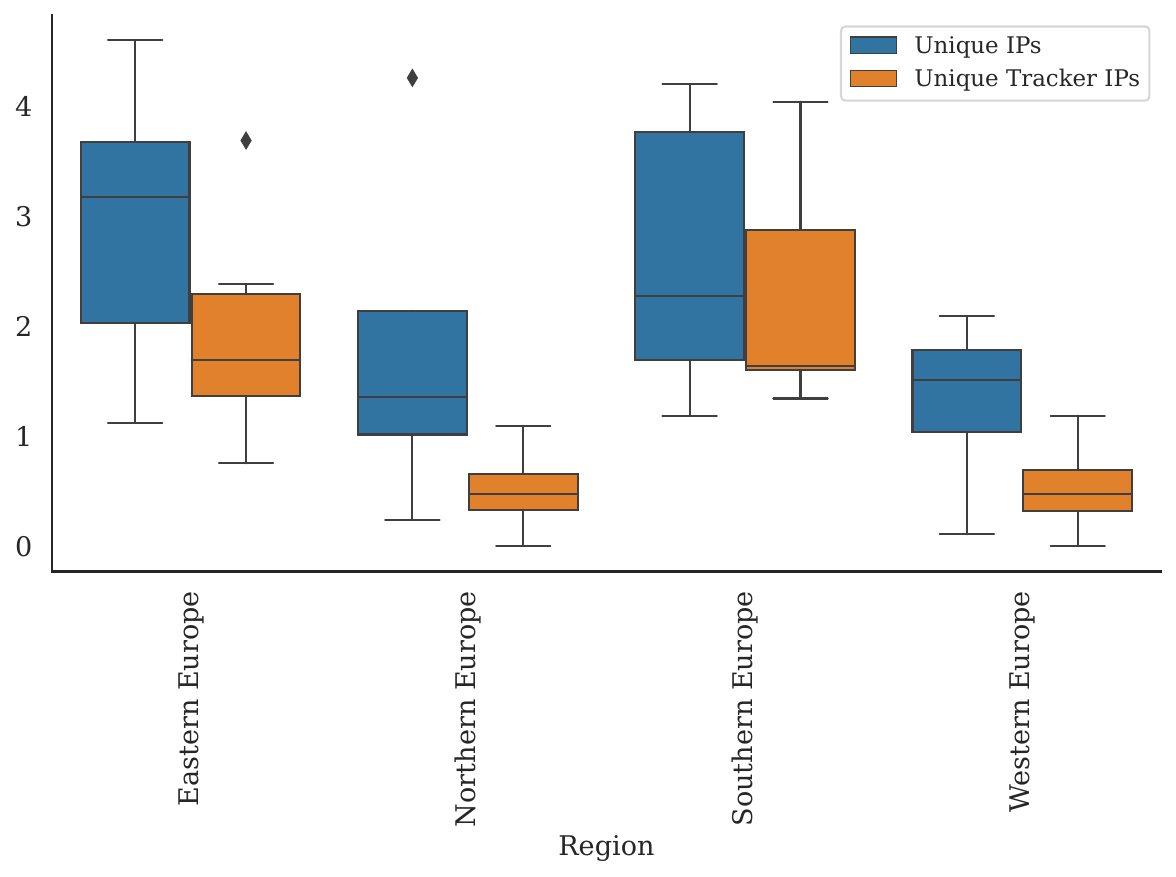}
    \caption{Boxplots showing the percentage of non-adequate server IPs and tracker IPs for the countries 
present in each EU region.}
\label{fig:regional}
\end{figure}

The findings are shown in Fig.~\ref{fig:regional}. The rates
of presence in non-adequate countries is higher in Southern and Eastern 
Europe, compared with Northern and Western Europe, for both server IPs and tracker IPs. 
We use an ANOVA test to determine whether these regional differences are statistically significant.
For server IPs, they are not ($0.25$). For tracker IPs, however, the difference is significant ($0.013$).
This latter category presents bigger privacy risks.
This disparity can lead to higher risks of privacy harms for users in Eastern and Southern Europe.

\subsection{Cookies}
In this subsection, we present an analysis of the cookies that were
loaded by initial sites that contacted at least one server in a non-adequate country.
We find substantial evidence that servers in non-adequate countries 
are engaging in user tracking activities.

Of the 1,233 non-adequate instances observed in \S~\ref{subsec:finalsample}
(recall that an instance is an initial site loaded from an ASCP),
we find that 
800 retrieve and store non-empty cookies. 
Of these, 236 websites load cookies with unique identifiers, for a total of 9,615 cookies. 
Unique identifiers pose a potential privacy harm, as they can be used to track
users beyond the website they are currently browsing.
These cookies contain 1,150 unique identifiers. 
The unique identifiers~\cite{munir2023cookiegraph} in a cookie's name or value 
can provide valuable context about the organization that issued or uses the cookie. 
We also used Cookiedatabase~\cite{cookiedatabase} for identifying organizations based on the unique identifier.
We found 480 cookies that contain an identifer $\texttt{\_ga}$, 
which indicates they were set by Google Analytics. 
Similarly, 443 cookies have $\texttt{\_gid}$ and 234 have $\texttt{\_\_gfp\_64b}$, which indicates 
they are used by Google services for analytics and ad personalization (DoubleClick), respectively.

\begin{center}
\begin{figure}
    \centering
    \includegraphics[width=\linewidth]{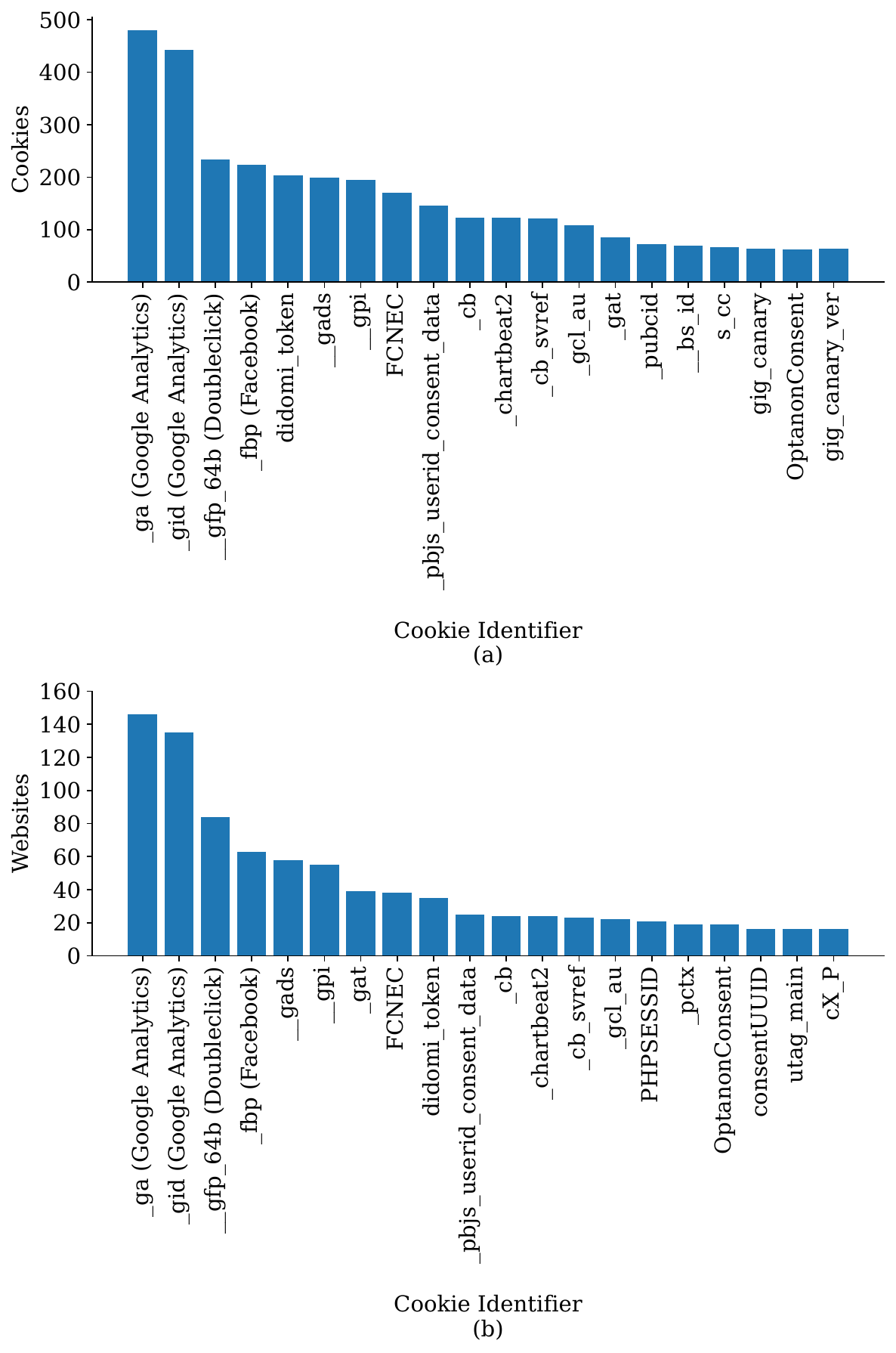}
    \caption{(a) Cookie identifier associated with the number of cookies. (b) Cookie identifier and number of website they are embedded.}
\label{fig:cookie_website}
\end{figure}
\end{center}

Other popular cookie identifier were, $\texttt{\_fbp}$ (223 cookies), 
set by Facebook for marketing. Additionally, we observed various identifiers 
that are related to consent management such as \path{\_pbjs\_userid\_consent\_data} and $\texttt{OptanonConsent}$.  
Consent management ensures compliance with privacy regulations by enabling users to control their data collection preferences.
Figure ~\ref{fig:cookie_website}(a) illustrates the distribution of the most frequently observed cookies.

Our analysis also identified the most commonly occurring cookies across websites. 
The most frequent cookie was $\texttt{\_ga}$ (Google Analytics), found on 146 websites, 
followed by $\texttt{\_gid}$ (Google Analytics) on 135 websites, $\texttt{\_\_gfp\_64b}$ (Doubleclick) 
on 84 and $\texttt{\_fbp}$ (Facebook) on 63 websites. 
Figures~\ref{fig:cookie_website}(b) illustrates and highlights the prevalence of tracking and analytics cookies across websites.

%% file: related.tex
\section{Discussion}
In this section we discuss our findings on data localization compliance
in the EU.

\subsection{Causes for Servers/Trackers in Non-Adequate Destinations}
Our method does not reveal the causes for web servers or trackers to be located in 
non-adequate countries, which is potentially unlawful in the EU.
We speculate about some potential causes here.
First, the content provider is intent in serving EU users from the EU (or an adequate country),
but in the presence of temporary server or router outages, it directs the user request to a backup
location elsewhere.
Second, the content owners may not know that the user is located in the EU,
as Internet protocols were not designed with physical nor jurisdictional constraints.
For instance, \textit{vk.com} might mistakenly infer that a user is located within Russia, thus
directing them to their servers there.
Third, in some instances, performance considerations may trump legal compliance.
For instance, users in Romania might experience much higher latency if their
content is served from a server in, say, Belgium, rather than Turkey, which is much closer.

\subsection{US as Destination}
The US is the most frequent non-adequate destination in our data,
observed from all but two EU countries in the sample.
The US is, of course, a key provider of cloud services,
and the home base of most of the world's largest content providers.
Partly given the amount of economic activity spurred by data transfers
between the US and the EU, the executive branches of both jurisdictions are intent on generating a
framework that broadly authorizes such transfers~\cite{EUUSData30:online,FACTSHEE89:online}. 
Such a framework is not guaranteed to stand up in the EU courts as previous attempts at
authorizing transfers from the EU to the US have failed in the EU's legal system~\cite{ThirdTim84:online}.
Whether the latest
EU-US framework will pass muster under future judicial review remains an open question
according to legal experts~\cite{ceps}.

\subsection{Regulatory Considerations}
The wide variety of trackers observed, loaded from an also varied
set of popular sites in each country, may pose challenges of scale for case-by-case regimes
such as the Standard Contractual Clauses (SCCs)~\cite{sccs},
particularly when auditing compliance. 
These have been applied to US companies~\cite{leglobal}
and are also in consideration~\cite{sccs} for data transfers to countries in 
Southeast Asia (ASEAN). We observed multiple ASEAN member countries~\cite{asean}
as destinations in our data: Thailand, Singapore and Malaysia. While
they are relatively uncommon destinations in our sample, EU-ASEAN
agreements~\cite{sccs} might change that in the future.

\subsection{News Sites}
The concentration we observed, where trackers in non-adequate countries
are primarily loaded by news sites, provides a potential opportunity
when designing data localization auditing frameworks in other jurisdictions
outside the EU: the websites in this
category should be evaluated first. This prioritization might 
be particularly useful in pilot evaluations of compliance with newly introduced
data localization requirements, or in environments where data collection is constrained
for instance by available bandwidth or where electric supply is less reliable.

\subsection{Regional Differences}
Perhaps the most surprising result in our study is the
differing rates of compliance across EU regions. While
certainly there are cultural and economic differences between
them, in theory it is not more challenging for a content provider
to deliver content from within Europe: there is ample available infrastructure
within the continent. However the disparities in 
compliance with data localization principles for trackers were significant. 
This uneven compliance with regulation intended to 
protect user privacy poses questions in equity and fairness:
the wealthier regions of the EU (Northern and Western) might 
be able to provide more uniform protections to their users than the 
less affluent regions in the south and east. Extrapolating this to
other jurisdictions who may be considering, or have recently implemented
data localization requirements, these equity and fairness considerations
are worth examining empirically. Empirical audits as the one we have presented in 
this study can reveal these issues 
with the greatest clarity.

\subsection{Russia and Finland}
The case of \textit{vk.com}, as accessed in Finland, presents a particular challenge
for data privacy. The company's CEO has been sanctioned by the EU in connection 
with the Ukraine conflict~\cite{vksanction}; the company 
is also state-owned (by Russia)~\cite{russiavk}. However,
since the services provided by VK are popular among Russian speakers,
which are a substantial portion of the population of many Eastern EU countries
and Finland, an outright ban of the service would negatively impact
these groups. We do find that \textit{vk.com} loads third party trackers in Russia,
creating a potential geopolitical challenge in addition to the usual potential privacy harms.

\input{testbed-test.tex}

\section{Proxy Location Validation}
\label{sec:proxyval}
We conduct an experiment to investigate whether
BrightData's claims about requests being routed through 
an AS-Country Pair are accurate. To this
end, we set up a web server at Northeastern University and
send HTTP requests through BrightData from each 
ASCP. All of the requests this server received were IPv4,
and we take steps to preserve the privacy of BrightData users
(who host the proxies in their own devices) by recording only the 
/24 subnet from which we received the request to our university server. 
We then compare the country and AS claimed by 
BrightData with those identified by geolocation database 
Maxmind~\cite{maxmindgeoloc}.
We fetch the AS and country for every IP in the /24 prefix through Maxmind.

We find that BrightData seems to be almost always routing
requests through the ASCP they claim.
Of the 2,319 valid requests received by this server from BrightData,
all but five are accurate. Thus, 2,314 requests have an IP that is 
part of a /24 prefix entirely present in the same ASCP according
to Maxmind.
The five exceptions include two where the country does not match (but the AS does), two where the AS does not match
(but the country does), and one where neither AS nor country are a match.
Therefore, we conclude that BrightData is an appropriate proxy to use for the 
purposes of routing requests through a specific AS in a given country.

We acknowledge that geolocation databases are prone to errors. However,
since we are working at the country level granularity, 
these errors are less common~\cite{10.1007/978-3-030-98785-5_6,10.1145/1971162.1971171}.
Of course, it is possible that both BrightData and Maxmind are often both incorrect and in agreement about
the ASCP where a user is located, but we argue that this is a remote possibility.

\section{Limitations}
In this section, we describe the main limitations of our approach. 

\subsection{Domain Exclusions}
\label{sec:domexc}
As described in \S~\ref{sec:domains}, our framework excludes
both Google-owned and adult websites as initial domains.
This exclusion is caused by BrightData rules. Thus,
our method is not able to study
these two groups of domains, particularly as
initial sites. For Google domains, this limitation
does not apply when they are loaded as third-parties by
other websites that are not Google owned.

\subsection{Potential for False Positives}
\label{sec:falsepos}
Since we do not have access to ground truth 
on physical server location, the rate of false positives in our
results is not known. 
For instance, there could be interactions between the 
server response rate to traceroutes and their geographic
location, which would impact our findings.
While we conducted a validation experiment
using servers with known location (\S~\ref{sec:testbed}),
it is still possible that the servers loaded by popular EU sites,
\textit{i.e.}, the servers we study in our main results section,
respond differently to our measurements than the servers in the
validation experiments. Furthermore, 
the EU servers in the validation experiment may treat 
traceroutes differently from other traffic,
potentially impacting our latency-based geolocation techniques.
Finally, the scale of our validation experiment is smaller than
the experimental setup in our main results,
which may impact the former's generalizability.
Therefore, the false positive rates in the validation 
experiment %
might differ from those in our main results.
Due to all the aforementioned factors, 
the lack of information regarding false positive rates and
precision in our main results is a limitation of this study.

\subsection{Alternative CDN Nodes}
A key limitation of our study 
is that we do not collect alternative server locations
for each destination, for instance available CDN nodes
~\cite{10.1145/1159913.1159962} from each ASCP.
This additional CDN-node data
would  potentially reveal whether the content providers are making their best effort to
comply with GDPR's data localization policies, \textit{i.e.}, they are
picking compliant servers even when they offer worse performance.
Alternatively, the content providers could instead primarily be selecting these servers
based on performance attributes such as latency.
Thus, our study is not able to distinguish between 
server placement that is based on performance constraints or
GDPR data localization compliance.

\subsection{Inference of GDPR Violations}

Our study is not able to make final determinations on GDPR violations. 
This is because legal exemptions may exist to mitigate the instances
where we have identified servers in non-adequate countries.
Further, a ruling on any potential GDPR violation would require additional
context on the specific data that was transferred and any legal contracts
in place between various entities, including privacy regulators~\cite{sccs}.
Ultimately, as such a determination of a GDPR violation would likely 
take place in a judicial context and be subject
to considerations beyond the physical location of a server. 

\subsection{Framework Applicability in Other Regions}
Our framework has limited applicability beyond the EU due to measurement
infrastructure density. Previous work has found that 
RIPE Atlas has more density of
deployment in Europe compared to other regions~\cite{ripeatlasbias,10198985}.
While geographic bias in the BrightData platform has not been as rigurously
quantified, even a cursory look at their top proxy locations reveals potential bias,
with the US and India having a similar number of IPs available even though the
latter is considerably larger in terms of Internet users.~\cite{topproxylocations}

\section{Related Work}
\label{sec:related}

Iordanou et al. ~\cite{10.1145/3278532.3278561} studies the
cross-border web tracking that targets users in the EU,
specifically the geographic scope of these data flows.
There are some similarities between this study and our work.
As in our study,
the previous work leverages large-scale Internet measurements,
partially launched from real end-user devices,
to study the location of servers responding to requests in the EU.
Both the related work and our study conclude that 
most traffic stays within the EU (or, in our case,
in the EU and adequate third countries); and
both also rely on RIPE IPmap for server geolocation, though our
framework also collects both source- and destination-based
measurements, and rDNS data, to validate each individual server inferred to be 
in a non-adequate country.

However, this previous work ~\cite{10.1145/3278532.3278561} differs from ours
in three key ways.
First, their focus is entirely on 
tracking servers, as they are interested in tracking flows, whereas we
investigate both general-purpose servers as well as tracking servers,
as we are instead interested in auditing data localization.
Second, the related work's
focus is on maximizing coverage of potential \textit{tracking flows},
whereas our study is focused on obtaining a representative
sample of \textit{source networks} in the EU. Third,
Iordanou et al. launch measurements from a browser
extension deployed to 183 EU users over a longer
period of several months, and intersect the
tracking servers there observed in ISP NetFlow data from
four large networks in three countries, whereas we rely 
on a point-in-time collection from a proxy deployed to more than 1,000 networks
in 19 countries.

Previous work has also tackled the issue of identifying cross-border data 
transfers. %
Guam{\'a}n et al.~\cite{9328756} study
the geographic spread of data flows originating from Android apps and evaluates
whether these flows comply with both the developer's privacy policies as
well as the GDPR. 
Razaghpanah et al. ~\cite{DBLP:conf/ndss/RazaghpanahNVSA18}
analyze tracking by mobile apps and whether these flows comply with
EU regulations, including an analysis of non-EU server locations.
Similarly, Nan et al. ~\cite{285453}
apply static analysis techniques to IoT companion apps (the smartphone apps needed to operate 
most IoT devices) and reveal data exposure caused by these devices.
Finally, Urban et al. ~\cite{10.1145/3366423.3380203} study third-party services in the
Web to generate a tree of dependencies that represents which additional services
are loaded by each initial third party contacted by a website.
While not a central focus of the latter two studies,
the authors do investigate the country where third-party servers that receive IoT/Web data (respectively) are located.
We note that all the studies cited in this paragraph rely on a geolocation
database (prone to inaccuracies ~\cite{10.1145/1971162.1971171, 10.1007/978-3-030-98785-5_6}) to pin the location of servers,
a method which often overstates the prevalence of servers located in the US~\cite{10.1145/3278532.3278561}, 
without conducting additional verification using Internet measurements,
as we have done in our work. Thus, we argue that our framework is more applicable
for auditing compliance with data localization requirements.

\section{Conclusion and Future Work}
A key component of the GDPR is the data localization regulation.
However, to date there was not a method to audit compliance
with this requirement at continental scale.
Our method fills this gap and provides a framework 
for empirically auditing compliance
with data localization principles and laws, and specifically investigate
such compliance for known trackers, which pose a greater privacy risk.
To accomplish this, we collect both browser-based data, the websites and 
domains loaded while browsing popular sites, and network-based data,
the servers that respond to such requests. We find that 
data localization requirements are broadly complied with in the EU,
though there are meaningful exceptions, which are more prevalent 
some regions. %

In the future, we plan on investigating the causes of high compliance rates,
disambiguating between legal compliance and performance prioritization by
web content companies. We further plan on applying this framework to regions
beyond the EU. In these regions, there are regulations that differ from the GDPR
on data localization to varying extents, providing opportunities for comparative
analyses of policy effectiveness with regards to data localization.

\section{Acknowledgements}
We thank the anonymous reviewers and the revision editor for their
valuable feedback. We are grateful to BrightData and RIPE Atlas
for allowing us to collect data using their platforms. 
This research was funded in part by the 
US National Science Foundation (NSF), Grants No. CNS 1955227 and CNS 2402963.
Author Gamero-Garrido was supported in part by Northeastern University's 
Future Faculty Fellowship and the Ford Foundation Postdoctoral Fellowship.

%% file: testbed-test.tex
\section{Server Geolocation Validation}
\label{sec:testbed}

To validate our server geolocation method, we build an experiment 
based on two sets of server IPs with known location.
Each of these sets is based either in a non-adequate country, the 
US,~\footnote{At the time of our main data collection. The validation experiment 
was conducted in Feb. 2025, when the US had already received an adequacy decision.~\cite{latham}}
or an adequate country, an EU member state.
To conduct this validation experiment, we followed the four steps 
of the methodology defined in \S~\ref{sec:servergeolocation} to perform 
server IP geolocation: RIPE IPmap geolocation, source-based constraints, 
destination-based constraints, and reverse DNS (rDNS).

The results of this experiment are shown in Tab.~\ref{tab:testbedgeo}.
In the following subsections, we describe this validation experiment.
We also infer rates of true/false positives and precision of our method.
Note that, as discussed in the limitations section (\S~\ref{sec:falsepos}), the 
rate of false positives might be different in our main results.

\subsection{Server Testbed in the US}
We have access 
to a US testbed, CloudLab ~\cite{CloudLab}, and thus to ground truth mapping between server IP and country of operation.
We conducted experiments using 200 such servers from this testbed that are
distributed across two different locations (L1 and L2), both in the US.
These servers are the destination IPs for this experiment. 
Further, we selected five source countries from 
the European Union (EU). We included the largest EU country, France (by land area), 
and four additional countries chosen randomly to represent different regions of the EU: 
Poland from Eastern Europe, Germany from Western Europe, Ireland from Northern Europe, 
and Spain from Southern Europe.
For each source country, we randomly selected 40 server IPs as destinations across 
the two testbed locations (L1 and L2). This results in a total of 200 destination IP-source
country pairs (DSCPs). 

The upper half of Tab.~\ref{tab:testbedgeo} shows the results of this
experiment. In the rest of this paragraph, we describe these 
findings in more detail. 
We found that RIPE IPmap did not provide country-level geolocation 
for three server IPs. %
In the source-based traceroutes, five server IPs failed to respond, and 20 did not meet the 90\% 
latency threshold described in \S~\ref{sec:source-based}. 
We further excluded two server IPs
using destination-based constraints. %
Finally, when we analyzed rDNS data, 
three of these servers had clues indicating they were not located in the city identified by RIPE IPmap,
but they were still located in the US. As our method is concerned with
country-level geolocation, we keep the server geolocation unchnaged from previous steps.
In summary, our final dataset correctly geolocated 170 DSCPs 
as being in the US,
and discarded 30 DSCPs due to the constraints described above.

\subsection{AWS Servers in the EU}

To validate our method for adequate server IPs, 
we conducted an experiment using 1,000 IPs in the EU 
advertised by Amazon Web Services~\cite{aws_ip_ranges}. 
We selected these IPs by intersecting the AWS-published IP ranges
with the ISI IP Hitlist,~\cite{ant_hitlist} which estimates the likelihood
that a server will respond to network measurements.
The IPs we select have a score of 99 on the Hitlist. 
We take this step in an attempt to maximize response rates to our measurements 
since, unlike on the US testbed, 
we have no direct knowledge of whether the AWS IPs are currently
both routed and in use. This is also why we increase the 
number of server IPs in the EU relative to the US experiments in the previous
subsection.

Similarly to our treatment of the US testbed, we divide the server IPs 
into five groups corresponding to the same five 
EU source countries. 
The lower half of Tab.~\ref{tab:testbedgeo} shows the results of this experiment.

From these 1,000 DSCPs, we find that RIPE IPMap 
does not assign a country in 36 cases, and erroneously geolocates 
270 DSCPs outside of the EU in non-adequate countries: 
one in Pakistan, the rest in the US. 
After conducting both our source- and destination-based measurements, 
and applying speed of light constraints, these 270 DSCPs are all discarded.
At this stage, 27 DSCPs are correctly labeled as being in an (adequate)
EU country.
rDNS confirms this assertion in all 27 cases.

\begin{table}
  \caption{DSCP Geolocation for server validation experiment. 
ST/DT: Source/Destination Traceroutes. 
TB: Server Testbed. *No hostname/no geohint.}
  \label{tab:testbedgeo}
  \begin{tabular}{ccccc}
    \toprule
    \textbf{Method} & \textbf{DSCPs} & \textbf{Unres-} & \textbf{Adequate} & \textbf{Non-}\\
     &  & \textbf{ponsive} &  & \textbf{adequate} \\
    \toprule
    ST: US TB & 197  & 5 & 0 & 172 \\
    DT: US TB & 172 & 0 & 0& 170 \\
    rDNS: US TB & 170 & 99/0* & 0 & 170 \\
    \midrule
    ST: AWS & 964  & 910 & 50  & 0 \\
    DT: AWS & 50  & 21 & 27  & 0 \\
    rDNS: AWS & 27  & 0/0* & 27  & 0 \\
  \bottomrule
\end{tabular}
\end{table}

\subsection{True/False Positives and Negatives, Precision}
In this subsection, we describe the rates of
true/false positives and negatives,
along with precision, that result
from our validation experiments.

\textbf{False positives and true negatives}. 
The outcome of the AWS experiment is either a false positive,
where a server 
in an adequate country is incorrectly inferred as being in a
non-adequate country; or a true negative,
where a server in an adequate country is correctly labeled as being so.

Given the results of the AWS validation experiment, 
we infer \textit{no false positives}.
This is because all servers initially labeled as being outside the 
EU by RIPE IPMap were correctly discarded---assigned a negative label---by 
subsequent steps in our method. 
Note that our source- and destination-based filters 
aggressively discard DSCPs because they are
either unresponsive or the latency fails our speed of light constraints.
Thus, after all filters are applied, 
the \textit{true negative} rate is 100\%.

In aggregate, these results suggest that the coverage of our method is limited,
because it excludes servers that may actually be located 
in a non-adequate country. 
Simultaneously, our method reduces the likelihood of false positives. 
Thus, in this experiment, our method is working as intended.

\textbf{True positives and false negatives}. Recall that the US was a 
non-adequate destination at the time of our main data collection,
so all servers in the US testbed
have a non-adequate country as their ground truth location; %
thus, all DSCPs in the US testbed validation experiment are either
correctly allocated to the US, a true positive, or discarded, a false negative.
Our method correctly identified 170 DSCPs as being in the US,
of 200 DSCPs known to be in the US.
Therefore, the \textit{true positive} rate of our method in this experiment is 85\%,
and the \textit{false negative} rate is 15\%.

\textbf{Precision}. From the aforementioned results,
in particular given the absence of false positives,
the inferred \textit{precision} of our method 
in the validation experiments is 1.0.

%% file: appendix.tex
\section*{A Latency Distributions}
\label{app:latency}
\begin{figure}
\vspace{-4mm}
    \centering
    \includegraphics[width=.46\textwidth]{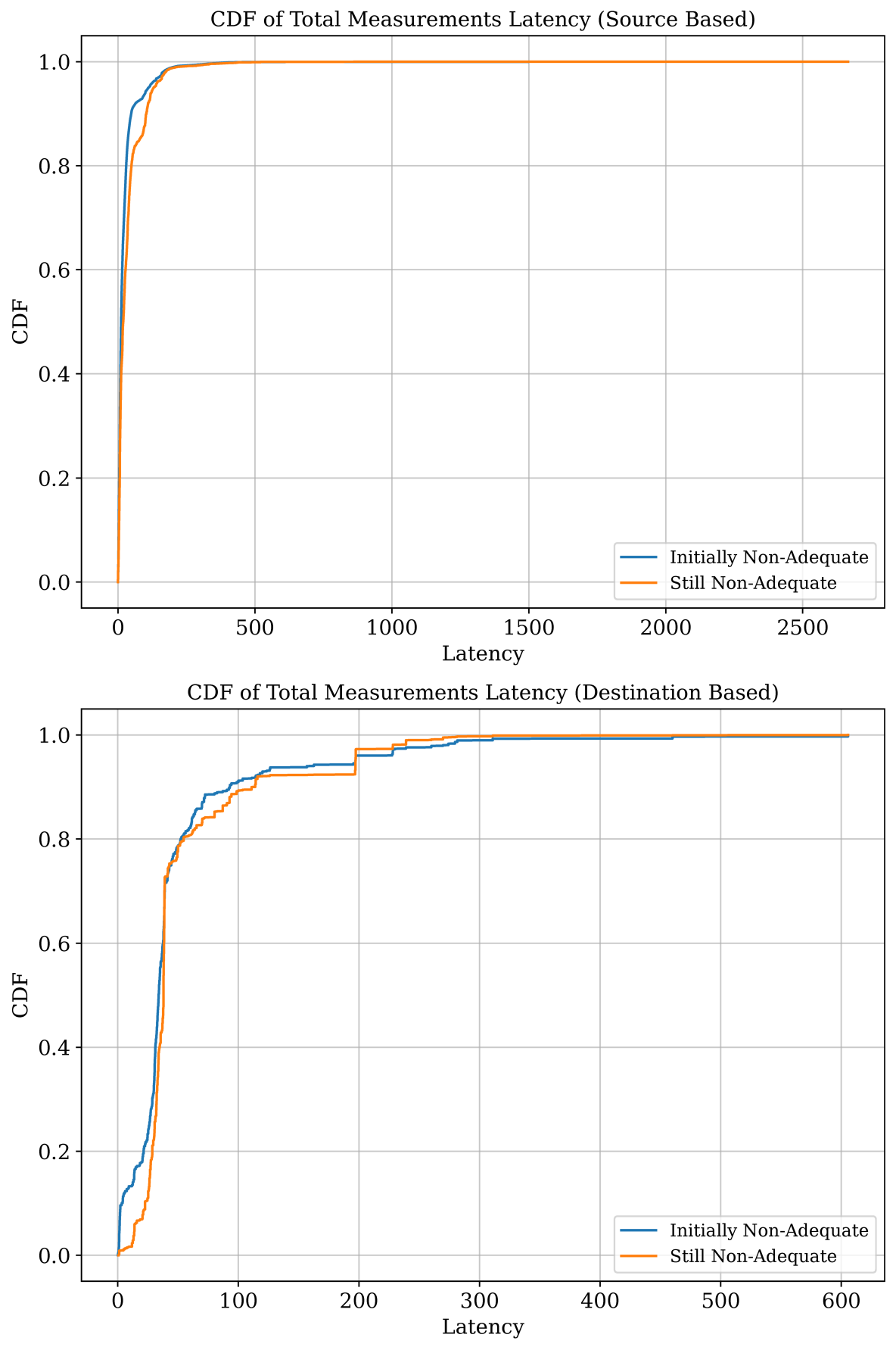}\\
    \caption{CDF of latency observed in source- and destination-based measurements.}
\label{fig:latencies}
\vspace{-4mm}
\end{figure}

In Fig.~\ref{fig:latencies}, we include the latency distribution from our measurements.
We include both source- and destination-based mesurements, both of which we use
(sequentially) to confirm the location of servers in non-adequate countries. The series
in the charts correspond to the latency observed to all candidate non-adequate servers
as well as those that we still label as non-adequate after the application of
speed of light constraints. As expected, the subset of servers that we confirm as being located in
non-adequate countries tend to have higher latencies than the larger group of initial candidate
servers.

\section*{B Ethics}
Our study does not collect personal information of any kind and does not
qualify as human subjects research. We collect data only from public, very popular web
sites and the resources that they load. We do not log in to any sites. We
use a separate browser that is routed through a proxy (not the user's own
browser). Thus we do not have access to any user's browsing history, data locally stored in
any user's device, any device/identifying information, nor any other private data.
The only exception is our limited experiment to confirm the accuracy of BrightData's
stated location of proxies; there, we only record the /24 subnet to which the device running the proxy
is connected.
In sum, our study is an investigation of the public web and does not pose significant ethical concerns.
The scale of our data collection, for all types of information we gather,
is unlikely to impact any services of the major websites we
study, and there are meaningful benefits in understanding compliance with online privacy regulations.

\section*{C Exclusion of the Netherlands}
\label{app:netherlands}
This appendix documents a minor correction made after the camera-ready version was published, involving one source country, the Netherlands, in our dataset. While the correction had no material impact on our findings, we include it here for transparency.

After peer review was complete and shortly after the paper was published on the PETS website~\cite{pets25}, we discovered a clerical error in our initial data collection. The error impacts a single source country, the Netherlands (NL). The initial sites for that country were erroneously based on those popular in Poland (PL) according to SimilarWeb. To correct this error, we have removed the Netherlands from our sample, reducing the number of EU countries we study by 5\% (from 20 to 19).

The impact of this error in our results is very limited for two reasons: first, the Netherlands did not feature prominently in our results, which were focused on the Eastern and Southern fringes of the EU; and second, because the domains we did collect from the Netherlands were a repeat of those we collected from Poland, the overall shape of our findings does not change.

Crucially, the destination-based measurements were not impacted because the targets were based on the intersection of candidate non-adequate IPs from the earlier source-based round, and all candidate non-adequate IPs that responded to NL requests were also observed in other EU countries. Most of the findings in the paper, including in the validation section, are similarly not impacted (except for the exclusion of the Netherlands).

We summarize the marginal changes we did observe in Tab.~\ref{tab:changes}.

\begin{table}[!t]
\centering
\caption{Summary of marginal changes due to correction}
\label{tab:changes}
\begin{tabular}{>{\raggedright\arraybackslash}p{7cm} >{\raggedright\arraybackslash}p{4cm} >{\raggedright\arraybackslash}p{4cm}}
\toprule
\textbf{Variable} & \textbf{Change due to error} & \textbf{Impact on results} \\
\midrule
Sample of countries and source networks (ASNs) & Source ASNs reduced from 1223 to 1132 & Exclusion of the Netherlands as a source country \\
\midrule
Average fraction of servers serving users in each EU country and located in non-adequate destination countries & Increase from 2.2\% to 2.3\% & None \\
\midrule
Number of source-based traceroutes launched towards candidate servers in non-adequate countries & Reduction from 9905 to 9592 & None \\
\midrule
Number of servers in non-adequate countries that are engaging in user tracking activities by loading cookies & Reduction from 824 to 800 & None \\
\midrule
P-value of the ANOVA test used to determine whether there is a significant difference in rates of non-adequate trackers across EU regions & Increase from 0.01 to 0.013 & None (the p-value is still well under 0.05) \\
\bottomrule
\end{tabular}
\end{table}

%% file: main.bbl

\providecommand{\noopsort}[1]{}
\begin{thebibliography}{63}

%%% the Web address.

\ifx \showCODEN    \undefined \def \showCODEN     #1{\unskip}     \fi
\ifx \showDOI      \undefined \def \showDOI       #1{#1}\fi
\ifx \showISBNx    \undefined \def \showISBNx     #1{\unskip}     \fi
\ifx \showISBNxiii \undefined \def \showISBNxiii  #1{\unskip}     \fi
\ifx \showISSN     \undefined \def \showISSN      #1{\unskip}     \fi
\ifx \showLCCN     \undefined \def \showLCCN      #1{\unskip}     \fi
\ifx \shownote     \undefined \def \shownote      #1{#1}          \fi
\ifx \showarticletitle \undefined \def \showarticletitle #1{#1}   \fi
\ifx \showURL      \undefined \def \showURL       {\relax}        \fi
% invisible to TeX
\providecommand\bibfield[2]{#2}
\providecommand\bibinfo[2]{#2}
\providecommand\natexlab[1]{#1}
\providecommand\showeprint[2][]{arXiv:#2}

\bibitem[ANT~Project(2024)]%
        {ant_hitlist}
\bibfield{author}{\bibinfo{person}{USC Information Sciences~Institute
  ANT~Project}.} \bibinfo{year}{2024}\natexlab{}.
\newblock \bibinfo{title}{IP Hitlist Dataset}.
\newblock
\newblock
\urldef\tempurl%
\url{https://ant.isi.edu/datasets/ip_hitlists/format.html}
\showURL{%
\tempurl}
\newblock
\shownote{Accessed: February 28, 2025}.


\bibitem[ASEAN(2024)]%
        {asean}
\bibfield{author}{\bibinfo{person}{ASEAN}.} \bibinfo{year}{2024}\natexlab{}.
\newblock \bibinfo{title}{ASEAN Member States}.
\newblock \bibinfo{howpublished}{\url{https://asean.org/member-states/}}.
\newblock


\bibitem[Atlas(2021)]%
        {RipeAtlasInfo}
\bibfield{author}{\bibinfo{person}{RIPE Atlas}.}
  \bibinfo{year}{2021}\natexlab{}.
\newblock \bibinfo{title}{Probe Archive}.
\newblock
  \bibinfo{howpublished}{\url{https://ftp.ripe.net/ripe/atlas/probes/archive/2021/12/20211101.json.bz2}}.
\newblock
\newblock
\shownote{(Data for November 2021.)}.


\bibitem[badmojr(2022)]%
        {GitHosts}
\bibfield{author}{\bibinfo{person}{badmojr}.} \bibinfo{year}{2022}\natexlab{}.
\newblock \bibinfo{title}{1Hosts}.
\newblock \bibinfo{howpublished}{\url{https://badmojr.gitlab.io/1hosts/}}.
\newblock
\newblock
\shownote{(Fetched in August 2022.)}.


\bibitem[Bell(2024)]%
        {russiavk}
\bibfield{author}{\bibinfo{person}{The Bell}.} \bibinfo{year}{2024}\natexlab{}.
\newblock \bibinfo{title}{Russia takes direct control of top social media
  networks}.
\newblock
  \bibinfo{howpublished}{\url{https://en.thebell.io/russia-takes-direct-control-of-top-social-media-networks/}}.
\newblock


\bibitem[Board(2020)]%
        {edpb2}
\bibfield{author}{\bibinfo{person}{European Data~Protection Board}.}
  \bibinfo{year}{2020}\natexlab{}.
\newblock \bibinfo{title}{Frequently Asked Questions on the judgment of the
  Court of Justice of the European Union in Case C-311/18 - Data Protection
  Commissioner v Facebook Ireland Ltd and Maximillian Schrems}.
\newblock
  \bibinfo{howpublished}{\url{https://www.edpb.europa.eu/sites/default/files/files/file1/20200724_edpb_faqoncjeuc31118_en.pdf}}.
\newblock


\bibitem[Board(2025)]%
        {edpb}
\bibfield{author}{\bibinfo{person}{European Data~Protection Board}.}
  \bibinfo{year}{2025}\natexlab{}.
\newblock \bibinfo{title}{International data transfers}.
\newblock
  \bibinfo{howpublished}{\url{https://www.edpb.europa.eu/sme-data-protection-guide/international-data-transfers_en}}.
\newblock


\bibitem[BrightData(2023a)]%
        {BrightData}
\bibfield{author}{\bibinfo{person}{BrightData}.}
  \bibinfo{year}{2023}\natexlab{a}.
\newblock \bibinfo{title}{BrightData}.
\newblock \bibinfo{howpublished}{\url{https://brightdata.com/}}.
\newblock


\bibitem[BrightData(2023b)]%
        {topproxylocations}
\bibfield{author}{\bibinfo{person}{BrightData}.}
  \bibinfo{year}{2023}\natexlab{b}.
\newblock \bibinfo{title}{Top Proxy Locations}.
\newblock \bibinfo{howpublished}{\url{https://brightdata.com/locations}}.
\newblock


\bibitem[Bui et~al\mbox{.}(2022)]%
        {bui2022opt}
\bibfield{author}{\bibinfo{person}{Duc Bui}, \bibinfo{person}{Brian Tang},
  {and} \bibinfo{person}{Kang~G Shin}.} \bibinfo{year}{2022}\natexlab{}.
\newblock \showarticletitle{Do opt-outs really opt me out?}. In
  \bibinfo{booktitle}{\emph{Proceedings of the 2022 ACM SIGSAC Conference on
  Computer and Communications Security}}. \bibinfo{pages}{425--439}.
\newblock


\bibitem[CEPS(2024)]%
        {ceps}
\bibfield{author}{\bibinfo{person}{CEPS}.} \bibinfo{year}{2024}\natexlab{}.
\newblock \bibinfo{title}{In the EU-US data transfer and privacy quarrel, the
  end is not in sight}.
\newblock
  \bibinfo{howpublished}{\url{https://www.ceps.eu/the-eu-us-data-transfers-and-privacy-quarrel-the-end-is-not-in-sight/}}.
\newblock


\bibitem[{CloudLab}(2025)]%
        {CloudLab}
\bibfield{author}{\bibinfo{person}{{CloudLab}}.}
  \bibinfo{year}{2025}\natexlab{}.
\newblock \bibinfo{title}{CloudLab}.
\newblock
\newblock
\urldef\tempurl%
\url{https://www.cloudlab.us/}
\showURL{%
\tempurl}


\bibitem[Commission(2022a)]%
        {Adequacy38:online}
\bibfield{author}{\bibinfo{person}{European Commission}.}
  \bibinfo{year}{2022}\natexlab{a}.
\newblock \bibinfo{title}{Adequacy decisions}.
\newblock
  \bibinfo{howpublished}{\url{https://commission.europa.eu/law/law-topic/data-protection/international-dimension-data-protection/adequacy-decisions_en}}.
\newblock
\newblock
\shownote{(Accessed on 02/20/2023)}.


\bibitem[Commission(2022b)]%
        {EUUSData30:online}
\bibfield{author}{\bibinfo{person}{European Commission}.}
  \bibinfo{year}{2022}\natexlab{b}.
\newblock \bibinfo{title}{EU-U.S. Data Privacy Framework, draft adequacy
  decision}.
\newblock
  \bibinfo{howpublished}{\url{https://ec.europa.eu/commission/presscorner/detail/en/qanda_22_7632}}.
\newblock
\newblock
\shownote{(Accessed on 02/20/2023)}.


\bibitem[Commission(2024)]%
        {sccs}
\bibfield{author}{\bibinfo{person}{European Commission}.}
  \bibinfo{year}{2024}\natexlab{}.
\newblock \bibinfo{title}{Standard Contractual Clauses (SCC)}.
\newblock
  \bibinfo{howpublished}{\url{https://commission.europa.eu/law/law-topic/data-protection/international-dimension-data-protection/standard-contractual-clauses-scc_en}}.
\newblock


\bibitem[Commission(2025)]%
        {ecreport}
\bibfield{author}{\bibinfo{person}{European Commission}.}
  \bibinfo{year}{2025}\natexlab{}.
\newblock \bibinfo{title}{Report on the first periodic review of the
  functioning of the adequacy decision on the EU-US Data Privacy Framework}.
\newblock
  \bibinfo{howpublished}{\url{https://commission.europa.eu/document/25695177-8073-4ce3-bf81-eb816dc6b468_en}}.
\newblock


\bibitem[Consulting(2025)]%
        {gdpreu}
\bibfield{author}{\bibinfo{person}{Intersoft Consulting}.}
  \bibinfo{year}{2025}\natexlab{}.
\newblock \bibinfo{title}{Art. 45 GDPR Transfers on the basis of an adequacy
  decision}.
\newblock \bibinfo{howpublished}{\url{https://gdpr-info.eu/art-45-gdpr/}}.
\newblock


\bibitem[{Cookie Database}(2025)]%
        {cookiedatabase}
\bibfield{author}{\bibinfo{person}{{Cookie Database}}.}
  \bibinfo{year}{2025}\natexlab{}.
\newblock \bibinfo{title}{Cookie Database}.
\newblock
\newblock
\urldef\tempurl%
\url{https://cookiedatabase.org/}
\showURL{%
\tempurl}


\bibitem[Council(2022)]%
        {Indiasd23:online}
\bibfield{author}{\bibinfo{person}{Atlantic Council}.}
  \bibinfo{year}{2022}\natexlab{}.
\newblock \bibinfo{title}{India's data localization pivot can revamp global
  digital diplomacy - Atlantic Council}.
\newblock
  \bibinfo{howpublished}{\url{https://www.atlanticcouncil.org/blogs/southasiasource/indias-data-localization-pivot/}}.
\newblock
\newblock
\shownote{(Accessed on 02/20/2023)}.


\bibitem[Crunch(2022)]%
        {Asitsdat63:online}
\bibfield{author}{\bibinfo{person}{Tech Crunch}.}
  \bibinfo{year}{2022}\natexlab{}.
\newblock \bibinfo{booktitle}{\emph{As its data flows woes grow, Google lobbies
  for quickie fix to EU-US transfers | TechCrunch}}.
\newblock Tech Crunch.
\newblock
\newblock
\shownote{(Accessed on 02/20/2023)}.


\bibitem[Cymru(2023a)]%
        {cymru}
\bibfield{author}{\bibinfo{person}{Team Cymru}.}
  \bibinfo{year}{2023}\natexlab{a}.
\newblock \bibinfo{title}{Whois}.
\newblock \bibinfo{howpublished}{\url{whois.cymru.com}}.
\newblock


\bibitem[Cymru(2023b)]%
        {astwoorg}
\bibfield{author}{\bibinfo{person}{Team Cymru}.}
  \bibinfo{year}{2023}\natexlab{b}.
\newblock \bibinfo{title}{Whois}.
\newblock \bibinfo{howpublished}{\url{whois.cymru.com}}.
\newblock


\bibitem[Division(2024)]%
        {uneu}
\bibfield{author}{\bibinfo{person}{United Nations~Statistics Division}.}
  \bibinfo{year}{2024}\natexlab{}.
\newblock \bibinfo{title}{Standard country or area codes for statistical use
  (M49)}.
\newblock
  \bibinfo{howpublished}{\url{https://unstats.un.org/unsd/methodology/m49/}}.
\newblock


\bibitem[EasyList(2022)]%
        {EasyList}
\bibfield{author}{\bibinfo{person}{EasyList}.} \bibinfo{year}{2022}\natexlab{}.
\newblock \bibinfo{title}{EasyList}.
\newblock \bibinfo{howpublished}{\url{https://easylist.to/}}.
\newblock
\newblock
\shownote{(Fetched in August 2022.)}.


\bibitem[{Gamero-Garrido, A. and Yu, K. and Shankar, S. V. and Singh, S. K. and
  Balasubramanian, S. and Wilcox, A. and Choffnes, D.}(2025)]%
        {githubrepo}
\bibfield{author}{\bibinfo{person}{{Gamero-Garrido, A. and Yu, K. and Shankar,
  S. V. and Singh, S. K. and Balasubramanian, S. and Wilcox, A. and Choffnes,
  D.}}} \bibinfo{year}{2025}\natexlab{}.
\newblock \bibinfo{title}{EU Data Localization Repository}.
\newblock
\newblock
\urldef\tempurl%
\url{https://github.com/such-in/EU-Data-Localization}
\showURL{%
\tempurl}


\bibitem[GDPRhub(2021)]%
        {DSBAustr4:online}
\bibfield{author}{\bibinfo{person}{GDPRhub}.} \bibinfo{year}{2021}\natexlab{}.
\newblock \bibinfo{title}{DSB (Austria) - 2021-0.586.257 (D155.027) - GDPRhub}.
\newblock
  \bibinfo{howpublished}{\url{https://gdprhub.eu/index.php?title=DSB_(Austria)\_-\_2021-0.586.257\_(D155.027)\#Further\_Resources}}.
\newblock
\newblock
\shownote{(Accessed on 02/20/2023)}.


\bibitem[Global(2024)]%
        {leglobal}
\bibfield{author}{\bibinfo{person}{L~\&~E Global}.}
  \bibinfo{year}{2024}\natexlab{}.
\newblock \bibinfo{title}{USA: New EU Standard Contractual Clauses – FAQs for
  U.S. Organisations}.
\newblock
  \bibinfo{howpublished}{\url{https://leglobal.law/2021/07/29/usa-new-eu-standard-contractual-clauses-faqs-for-u-s-organisations/}}.
\newblock


\bibitem[Guam{\'a}n et~al\mbox{.}(2021)]%
        {9328756}
\bibfield{author}{\bibinfo{person}{Danny~S. Guam{\'a}n},
  \bibinfo{person}{Jose~M. Del~Alamo}, {and} \bibinfo{person}{Julio~C. Caiza}.}
  \bibinfo{year}{2021}\natexlab{}.
\newblock \showarticletitle{GDPR Compliance Assessment for Cross-Border
  Personal Data Transfers in Android Apps}.
\newblock \bibinfo{journal}{\emph{IEEE Access}}  \bibinfo{volume}{9}
  (\bibinfo{year}{2021}), \bibinfo{pages}{15961--15982}.
\newblock
\urldef\tempurl%
\url{https://doi.org/10.1109/ACCESS.2021.3053130}
\showDOI{\tempurl}


\bibitem[House(2022)]%
        {FACTSHEE89:online}
\bibfield{author}{\bibinfo{person}{The~White House}.}
  \bibinfo{year}{2022}\natexlab{}.
\newblock \bibinfo{title}{FACT SHEET: President Biden Signs Executive Order to
  Implement the European Union-U.S. Data Privacy Framework - The White House}.
\newblock
  \bibinfo{howpublished}{\url{https://www.whitehouse.gov/briefing-room/statements-releases/2022/10/07/fact-sheet-president-biden-signs-executive-order-to-implement-the-european-union-u-s-data-privacy-framework/}}.
\newblock
\newblock
\shownote{(Accessed on 02/20/2023)}.


\bibitem[ICANN(2024)]%
        {whois}
\bibfield{author}{\bibinfo{person}{ICANN}.} \bibinfo{year}{2024}\natexlab{}.
\newblock \bibinfo{title}{Registration data lookup tool}.
\newblock \bibinfo{howpublished}{\url{https://lookup.icann.org/en}}.
\newblock


\bibitem[Iordanou et~al\mbox{.}(2018)]%
        {10.1145/3278532.3278561}
\bibfield{author}{\bibinfo{person}{Costas Iordanou}, \bibinfo{person}{Georgios
  Smaragdakis}, \bibinfo{person}{Ingmar Poese}, {and} \bibinfo{person}{Nikolaos
  Laoutaris}.} \bibinfo{year}{2018}\natexlab{}.
\newblock \showarticletitle{Tracing Cross Border Web Tracking}. In
  \bibinfo{booktitle}{\emph{Proceedings of the Internet Measurement Conference
  2018}} (Boston, MA, USA) \emph{(\bibinfo{series}{IMC '18})}.
  \bibinfo{publisher}{Association for Computing Machinery},
  \bibinfo{address}{New York, NY, USA}, \bibinfo{pages}{329–342}.
\newblock
\showISBNx{9781450356190}
\urldef\tempurl%
\url{https://doi.org/10.1145/3278532.3278561}
\showDOI{\tempurl}


\bibitem[Katz-Bassett et~al\mbox{.}(2006)]%
        {10.1145/1177080.1177090}
\bibfield{author}{\bibinfo{person}{Ethan Katz-Bassett},
  \bibinfo{person}{John~P. John}, \bibinfo{person}{Arvind Krishnamurthy},
  \bibinfo{person}{David Wetherall}, \bibinfo{person}{Thomas Anderson}, {and}
  \bibinfo{person}{Yatin Chawathe}.} \bibinfo{year}{2006}\natexlab{}.
\newblock \showarticletitle{Towards IP Geolocation Using Delay and Topology
  Measurements}. In \bibinfo{booktitle}{\emph{Proceedings of the 6th ACM
  SIGCOMM Conference on Internet Measurement}} (Rio de Janeriro, Brazil)
  \emph{(\bibinfo{series}{IMC '06})}. \bibinfo{publisher}{Association for
  Computing Machinery}, \bibinfo{address}{New York, NY, USA},
  \bibinfo{pages}{71---84}.
\newblock
\showISBNx{1595935614}
\urldef\tempurl%
\url{https://doi.org/10.1145/1177080.1177090}
\showDOI{\tempurl}


\bibitem[Katz-Bassett and Smaragdakis(2021)]%
        {10.1145/3402413.3402415}
\bibfield{author}{\bibinfo{person}{Ethan Katz-Bassett} {and}
  \bibinfo{person}{Georgios Smaragdakis}.} \bibinfo{year}{2021}\natexlab{}.
\newblock \showarticletitle{Seven Years in the Life of Hypergiants' off-Nets}.
\newblock  (\bibinfo{year}{2021}), \bibinfo{pages}{516--533}.
\newblock
\showISBNx{9781450383837}
\urldef\tempurl%
\url{https://doi.org/10.1145/3452296.3472928}
\showDOI{\tempurl}


\bibitem[Latham and Watkins(2023)]%
        {latham}
\bibfield{author}{\bibinfo{person}{Latham} {and} \bibinfo{person}{LLP
  Watkins}.} \bibinfo{year}{2023}\natexlab{}.
\newblock \bibinfo{title}{EU-US Data Privacy Framework Goes Live: What Are the
  Practical Implications?}
\newblock
  \bibinfo{howpublished}{\url{https://www.globalprivacyblog.com/2023/08/eu-us-data-privacy-framework-goes-live-what-are-the-practical-implications/}}.
\newblock


\bibitem[Le~Pochat et~al\mbox{.}(2023)]%
        {Tranco}
\bibfield{author}{\bibinfo{person}{Victor Le~Pochat}, \bibinfo{person}{Tom
  Van~Goethem}, \bibinfo{person}{Samaneh Tajalizadehkhoob},
  \bibinfo{person}{Maciej Korczy{\'n}ski}, {and} \bibinfo{person}{Wouter
  Joosen}.} \bibinfo{year}{2023}\natexlab{}.
\newblock \bibinfo{title}{A research-oriented top sites ranking hardened
  against manipulation - Tranco}.
\newblock \bibinfo{howpublished}{\url{https://tranco-list.eu/}}.
\newblock
\newblock
\shownote{(Accessed on 04/10/2024)}.


\bibitem[Lexology(2022)]%
        {ThirdTim84:online}
\bibfield{author}{\bibinfo{person}{Lexology}.} \bibinfo{year}{2022}\natexlab{}.
\newblock \bibinfo{title}{Third Time's a Charm? The New EU-US Data Privacy
  Framework and the US's Pursuit of an EU Adequacy Decision under GDPR -
  Lexology}.
\newblock
  \bibinfo{howpublished}{\url{https://www.lexology.com/library/detail.aspx?g=b395d1f6-00c6-4081-ae20-91de58e4c109}}.
\newblock
\newblock
\shownote{(Accessed on 02/20/2023)}.


\bibitem[Luckie et~al\mbox{.}(1 12)]%
        {geodns}
\bibfield{author}{\bibinfo{person}{M. Luckie}, \bibinfo{person}{B. Huffaker},
  \bibinfo{person}{A. Marder}, \bibinfo{person}{Z. Bischof},
  \bibinfo{person}{M. Fletcher}, {and} \bibinfo{person}{k. claffy}.}
  \bibinfo{year}{2021-12}\natexlab{}.
\newblock \showarticletitle{Learning to Extract Geographic Information from
  Internet Router Hostnames}. In \bibinfo{booktitle}{\emph{ACM SIGCOMM
  Conference on emerging Networking EXperiments and Technologies (CoNEXT)}}.
\newblock


\bibitem[Maxmind(2023)]%
        {maxmindgeoloc}
\bibfield{author}{\bibinfo{person}{Maxmind}.} \bibinfo{year}{2023}\natexlab{}.
\newblock \bibinfo{title}{Maxmind Geolocation Data}.
\newblock
  \bibinfo{howpublished}{\url{https://www.maxmind.com/en/geoip2-services-and-databases}}.
\newblock


\bibitem[McKinsey(2023)]%
        {Dataloca93:online}
\bibfield{author}{\bibinfo{person}{McKinsey}.} \bibinfo{year}{2023}\natexlab{}.
\newblock \bibinfo{title}{Data localization and new competitive opportunities |
  McKinsey | McKinsey}.
\newblock
  \bibinfo{howpublished}{\url{https://www.mckinsey.com/capabilities/risk-and-resilience/our-insights/localization-of-data-privacy-regulations-creates-competitive-opportunities}}.
\newblock
\newblock
\shownote{(Accessed on 02/20/2023)}.


\bibitem[Me(2024)]%
        {whotracksme}
\bibfield{author}{\bibinfo{person}{Who~Tracks Me}.}
  \bibinfo{year}{2024}\natexlab{}.
\newblock \bibinfo{title}{Who Tracks Me}.
\newblock \bibinfo{howpublished}{\url{https://www.ghostery.com/whotracksme}}.
\newblock


\bibitem[Moss(2024)]%
        {vksanction}
\bibfield{author}{\bibinfo{person}{Sebastian Moss}.}
  \bibinfo{year}{2024}\natexlab{}.
\newblock \bibinfo{title}{EU sanctions Rostelecom president, Yandex deputy CEO,
  and VK Company CEO}.
\newblock
  \bibinfo{howpublished}{\url{https://www.datacenterdynamics.com/en/news/eu-sanctions-rostelecom-president-yandex-deputy-ceo-and-vk-company-ceo/}}.
\newblock


\bibitem[Munir et~al\mbox{.}(2023)]%
        {munir2023cookiegraph}
\bibfield{author}{\bibinfo{person}{Shaoor Munir}, \bibinfo{person}{Sandra
  Siby}, \bibinfo{person}{Umar Iqbal}, \bibinfo{person}{Steven Englehardt},
  \bibinfo{person}{Zubair Shafiq}, {and} \bibinfo{person}{Carmela Troncoso}.}
  \bibinfo{year}{2023}\natexlab{}.
\newblock \showarticletitle{Cookiegraph: Understanding and detecting
  first-party tracking cookies}. In \bibinfo{booktitle}{\emph{Proceedings of
  the 2023 ACM SIGSAC Conference on Computer and Communications Security}}.
  \bibinfo{pages}{3490--3504}.
\newblock


\bibitem[Nan et~al\mbox{.}(2023)]%
        {285453}
\bibfield{author}{\bibinfo{person}{Yuhong Nan}, \bibinfo{person}{Xueqiang
  Wang}, \bibinfo{person}{Luyi Xing}, \bibinfo{person}{Xiaojing Liao},
  \bibinfo{person}{Ruoyu Wu}, \bibinfo{person}{Jianliang Wu},
  \bibinfo{person}{Yifan Zhang}, {and} \bibinfo{person}{XiaoFeng Wang}.}
  \bibinfo{year}{2023}\natexlab{}.
\newblock \showarticletitle{Are You Spying on Me? {Large-Scale} Analysis on
  {IoT} Data Exposure through Companion Apps}. In
  \bibinfo{booktitle}{\emph{32nd USENIX Security Symposium (USENIX Security
  23)}}. \bibinfo{publisher}{USENIX Association}, \bibinfo{address}{Anaheim,
  CA}, \bibinfo{pages}{6665--6682}.
\newblock
\showISBNx{978-1-939133-37-3}
\urldef\tempurl%
\url{https://www.usenix.org/conference/usenixsecurity23/presentation/nan}
\showURL{%
\tempurl}


\bibitem[Noyb(2020)]%
        {101Compl42:online}
\bibfield{author}{\bibinfo{person}{Noyb}.} \bibinfo{year}{2020}\natexlab{}.
\newblock \bibinfo{title}{101 Complaints on EU-US transfers filed}.
\newblock
  \bibinfo{howpublished}{\url{https://noyb.eu/en/101-complaints-eu-us-transfers-filed}}.
\newblock
\newblock
\shownote{(Accessed on 02/20/2023)}.


\bibitem[Paracha et~al\mbox{.}(2020)]%
        {paracha2020deeper}
\bibfield{author}{\bibinfo{person}{Muhammad~Talha Paracha},
  \bibinfo{person}{Balakrishnan Chandrasekara}, \bibinfo{person}{David
  Choffnes}, {and} \bibinfo{person}{Dave Levin}.}
  \bibinfo{year}{2020}\natexlab{}.
\newblock \showarticletitle{A Deeper Look at Web Content Availability and
  Consistency over HTTP/S}. In \bibinfo{booktitle}{\emph{2020 Network Traffic
  Measurement and Analysis Conference (TMA'20)}}.
\newblock


\bibitem[Parliament(2020)]%
        {schremsii}
\bibfield{author}{\bibinfo{person}{European Parliament}.}
  \bibinfo{year}{2020}\natexlab{}.
\newblock \bibinfo{title}{The CJEU judgment in the Schrems II case}.
\newblock
  \bibinfo{howpublished}{\url{https://www.europarl.europa.eu/RegData/etudes/ATAG/2020/652073/EPRS\_ATA(2020)652073\_EN.pdf}}.
\newblock


\bibitem[{Pavlos Sermpezis}(2022)]%
        {ripeatlasbias}
\bibfield{author}{\bibinfo{person}{{Pavlos Sermpezis}}.}
  \bibinfo{year}{2022}\natexlab{}.
\newblock \bibinfo{title}{Bias in Internet Measurement Infrastructure}.
\newblock
\newblock
\urldef\tempurl%
\url{https://labs.ripe.net/author/pavlos_sermpezis/bias-in-internet-measurement-infrastructure/}
\showURL{%
\tempurl}


\bibitem[{PETS 2025}(2025)]%
        {pets25}
\bibfield{author}{\bibinfo{person}{{PETS 2025}}.}
  \bibinfo{year}{2025}\natexlab{}.
\newblock \bibinfo{title}{PETS 2025}.
\newblock
\newblock
\urldef\tempurl%
\url{https://petsymposium.org/popets/2025/}
\showURL{%
\tempurl}


\bibitem[Poese et~al\mbox{.}(2011)]%
        {10.1145/1971162.1971171}
\bibfield{author}{\bibinfo{person}{Ingmar Poese}, \bibinfo{person}{Steve
  Uhlig}, \bibinfo{person}{Mohamed~Ali Kaafar}, \bibinfo{person}{Benoit
  Donnet}, {and} \bibinfo{person}{Bamba Gueye}.}
  \bibinfo{year}{2011}\natexlab{}.
\newblock \showarticletitle{IP geolocation databases: unreliable?}
\newblock \bibinfo{journal}{\emph{SIGCOMM Comput. Commun. Rev.}}
  \bibinfo{volume}{41}, \bibinfo{number}{2} (\bibinfo{date}{apr}
  \bibinfo{year}{2011}), \bibinfo{pages}{53---56}.
\newblock
\showISSN{0146-4833}
\urldef\tempurl%
\url{https://doi.org/10.1145/1971162.1971171}
\showDOI{\tempurl}


\bibitem[Razaghpanah et~al\mbox{.}(2018)]%
        {DBLP:conf/ndss/RazaghpanahNVSA18}
\bibfield{author}{\bibinfo{person}{Abbas Razaghpanah}, \bibinfo{person}{Rishab
  Nithyanand}, \bibinfo{person}{Narseo Vallina{-}Rodriguez},
  \bibinfo{person}{Srikanth Sundaresan}, \bibinfo{person}{Mark Allman},
  \bibinfo{person}{Christian Kreibich}, {and} \bibinfo{person}{Phillipa Gill}.}
  \bibinfo{year}{2018}\natexlab{}.
\newblock \showarticletitle{Apps, Trackers, Privacy, and Regulators: {A} Global
  Study of the Mobile Tracking Ecosystem}. In \bibinfo{booktitle}{\emph{25th
  Annual Network and Distributed System Security Symposium, {NDSS} 2018, San
  Diego, California, USA, February 18-21, 2018}}. \bibinfo{publisher}{The
  Internet Society}.
\newblock
\urldef\tempurl%
\url{https://www.ndss-symposium.org/wp-content/uploads/2018/02/ndss2018\_05B-3\_Razaghpanah\_paper.pdf}
\showURL{%
\tempurl}


\bibitem[RIPE(2024)]%
        {atlas}
\bibfield{author}{\bibinfo{person}{RIPE}.} \bibinfo{year}{2024}\natexlab{}.
\newblock \bibinfo{title}{{RIPE Atlas}}.
\newblock \bibinfo{howpublished}{\url{https://atlas.ripe.net/}}.
\newblock


\bibitem[Ruth et~al\mbox{.}(2022)]%
        {ruth2022world}
\bibfield{author}{\bibinfo{person}{Kimberly Ruth}, \bibinfo{person}{Aurore
  Fass}, \bibinfo{person}{Jonathan Azose}, \bibinfo{person}{Mark Pearson},
  \bibinfo{person}{Emma Thomas}, \bibinfo{person}{Caitlin Sadowski}, {and}
  \bibinfo{person}{Zakir Durumeric}.} \bibinfo{year}{2022}\natexlab{}.
\newblock \showarticletitle{A world wide view of browsing the world wide web}.
  In \bibinfo{booktitle}{\emph{Proceedings of the 22nd ACM Internet Measurement
  Conference}}. \bibinfo{pages}{317--336}.
\newblock


\bibitem[Saxon and Feamster(2022)]%
        {10.1007/978-3-030-98785-5_6}
\bibfield{author}{\bibinfo{person}{James Saxon} {and} \bibinfo{person}{Nick
  Feamster}.} \bibinfo{year}{2022}\natexlab{}.
\newblock \showarticletitle{GPS-Based Geolocation of Consumer IP Addresses}. In
  \bibinfo{booktitle}{\emph{Passive and Active Measurement: 23rd International
  Conference, PAM 2022, Virtual Event, March 28–30, 2022, Proceedings}}.
  \bibinfo{publisher}{Springer-Verlag}, \bibinfo{address}{Berlin, Heidelberg},
  \bibinfo{pages}{122---151}.
\newblock
\showISBNx{978-3-030-98784-8}
\urldef\tempurl%
\url{https://doi.org/10.1007/978-3-030-98785-5_6}
\showDOI{\tempurl}


\bibitem[Sermpezis et~al\mbox{.}(2023)]%
        {10198985}
\bibfield{author}{\bibinfo{person}{Pavlos Sermpezis}, \bibinfo{person}{Lars
  Prehn}, \bibinfo{person}{Sofia Kostoglou}, \bibinfo{person}{Marcel Flores},
  \bibinfo{person}{Athena Vakali}, {and} \bibinfo{person}{Emile Aben}.}
  \bibinfo{year}{2023}\natexlab{}.
\newblock \showarticletitle{Bias in Internet Measurement Platforms}. In
  \bibinfo{booktitle}{\emph{2023 7th Network Traffic Measurement and Analysis
  Conference (TMA)}}. \bibinfo{pages}{1--10}.
\newblock
\urldef\tempurl%
\url{https://doi.org/10.23919/TMA58422.2023.10198985}
\showDOI{\tempurl}


\bibitem[Services(2024)]%
        {aws_ip_ranges}
\bibfield{author}{\bibinfo{person}{Amazon~Web Services}.}
  \bibinfo{year}{2024}\natexlab{}.
\newblock \bibinfo{title}{AWS IP Address Ranges}.
\newblock
\newblock
\urldef\tempurl%
\url{https://ip-ranges.amazonaws.com/ip-ranges.json}
\showURL{%
\tempurl}
\newblock
\shownote{Accessed: February 28, 2025}.


\bibitem[SimilarWeb(2022)]%
        {SimilarWeb}
\bibfield{author}{\bibinfo{person}{SimilarWeb}.}
  \bibinfo{year}{2022}\natexlab{}.
\newblock \bibinfo{title}{Top Sites}.
\newblock
  \bibinfo{howpublished}{\url{https://www.similarweb.com/top-websites/germany/}}.
\newblock
\newblock
\shownote{(Data for July 2022.)}.


\bibitem[Su et~al\mbox{.}(2006)]%
        {10.1145/1159913.1159962}
\bibfield{author}{\bibinfo{person}{Ao-Jan Su}, \bibinfo{person}{David~R.
  Choffnes}, \bibinfo{person}{Aleksandar Kuzmanovic}, {and}
  \bibinfo{person}{Fabi\'{a}n~E. Bustamante}.} \bibinfo{year}{2006}\natexlab{}.
\newblock \showarticletitle{Drafting behind Akamai (travelocity-based
  detouring)}. In \bibinfo{booktitle}{\emph{Proceedings of the 2006 Conference
  on Applications, Technologies, Architectures, and Protocols for Computer
  Communications}} (Pisa, Italy) \emph{(\bibinfo{series}{SIGCOMM '06})}.
  \bibinfo{publisher}{Association for Computing Machinery},
  \bibinfo{address}{New York, NY, USA}, \bibinfo{pages}{435---446}.
\newblock
\showISBNx{1595933085}
\urldef\tempurl%
\url{https://doi.org/10.1145/1159913.1159962}
\showDOI{\tempurl}


\bibitem[Tessian(2022)]%
        {30Bigges7:online}
\bibfield{author}{\bibinfo{person}{Tessian}.} \bibinfo{year}{2022}\natexlab{}.
\newblock \bibinfo{booktitle}{\emph{30 Biggest GDPR Fines To-Date | Latest GDPR
  Fines | Updated 2022 | Tessian}}.
\newblock Thessian.
\newblock
\newblock
\shownote{(Accessed on 02/20/2023)}.


\bibitem[Union(2015)]%
        {schremsi}
\bibfield{author}{\bibinfo{person}{European Union}.}
  \bibinfo{year}{2015}\natexlab{}.
\newblock \bibinfo{title}{Judgment of the Court (Grand Chamber) of 6 October
  2015. Maximillian Schrems v Data Protection Commissioner.}
\newblock
  \bibinfo{howpublished}{https://eur-lex.europa.eu/legal-content/en/TXT/?uri=CELEX\%3A62014CJ0362}.
\newblock


\bibitem[Urban et~al\mbox{.}(2020)]%
        {10.1145/3366423.3380203}
\bibfield{author}{\bibinfo{person}{Tobias Urban}, \bibinfo{person}{Martin
  Degeling}, \bibinfo{person}{Thorsten Holz}, {and} \bibinfo{person}{Norbert
  Pohlmann}.} \bibinfo{year}{2020}\natexlab{}.
\newblock \showarticletitle{Beyond the Front Page:Measuring Third Party
  Dynamics in the Field}. In \bibinfo{booktitle}{\emph{Proceedings of The Web
  Conference 2020}} (Taipei, Taiwan) \emph{(\bibinfo{series}{WWW '20})}.
  \bibinfo{publisher}{Association for Computing Machinery},
  \bibinfo{address}{New York, NY, USA}, \bibinfo{pages}{1275–1286}.
\newblock
\showISBNx{9781450370233}
\urldef\tempurl%
\url{https://doi.org/10.1145/3366423.3380203}
\showDOI{\tempurl}


\bibitem[Verizon(2022)]%
        {MonthlyI86:online}
\bibfield{author}{\bibinfo{person}{Verizon}.} \bibinfo{year}{2022}\natexlab{}.
\newblock \bibinfo{title}{Monthly IP Latency Data Verizon Enterprise
  Solutions}.
\newblock
  \bibinfo{howpublished}{\url{https://www.verizon.com/business/terms/latency/}}.
\newblock
\newblock
\shownote{(Data for August 2022.)}.


\bibitem[WonderNetwork(2022)]%
        {GlobalPi56:online}
\bibfield{author}{\bibinfo{person}{WonderNetwork}.}
  \bibinfo{year}{2022}\natexlab{}.
\newblock \bibinfo{title}{Global Ping Statistics - WonderNetwork}.
\newblock \bibinfo{howpublished}{\url{https://wondernetwork.com/pings}}.
\newblock
\newblock
\shownote{(Data for November 2022.)}.


\bibitem[Zheutlin et~al\mbox{.}(2021)]%
        {zheutlin2021data}
\bibfield{author}{\bibinfo{person}{Alexander~R Zheutlin},
  \bibinfo{person}{Joshua~D Niforatos}, {and} \bibinfo{person}{Jeremy~B
  Sussman}.} \bibinfo{year}{2021}\natexlab{}.
\newblock \showarticletitle{Data-tracking on government, non-profit, and
  commercial health-related websites}.
\newblock \bibinfo{journal}{\emph{Journal of general internal medicine}}
  (\bibinfo{year}{2021}), \bibinfo{pages}{1--3}.
\newblock


\end{thebibliography}
